\documentclass[useAMS,usenatbib,usegraphicx]{mn2e}
\usepackage{bm}
\usepackage{times}

\title[Extinction curve and metal depletion]
{Effects of grain growth mechanisms on the extinction curve
and the metal depletion in the interstellar medium}
\author[Hirashita]{Hiroyuki Hirashita$^1$\thanks{E-mail:
    hirashita@asiaa.sinica.edu.tw}
and Nikolai V. Voshchinnikov$^2$\\
$^1$Institute of Astronomy and Astrophysics, Academia Sinica,
P.O. Box 23-141, Taipei 10617, Taiwan\\
$^2$Sobolev Astronomical Institute, St. Petersburg University,
Universitetskii prosp., 28, St. Petersburg 198504, Russia 
}
\date{2013 October 17}

\pagerange{\pageref{firstpage}--\pageref{lastpage}} \pubyear{2012}

\begin{document}
\label{firstpage}
\maketitle

\begin{abstract}
Dust grains grow their sizes in the interstellar clouds
(especially in molecular clouds) by accretion and
coagulation. Here we model and test these processes
by examining the consistency with the observed variation of
the extinction curves in the Milky Way. We find that, if we
simply use the parameters used in previous studies,
the model fails to explain the flattening of far-UV extinction
curve for large $R_V$ (flatness of optical extinction curve)
and the existence of carbon bump even in flat extinction
curves. This discrepancy is resolved by adopting
a `tuned' model, in which coagulation of carbonaceous dust
is less efficient (by a factor of 2) and that of silicate is
more efficient with the coagulation threshold removed.
The tuned model is also consistent with the relation between
silicon depletion
(indicator of accretion) and $R_V$
{if the duration of accretion and coagulation
is $\ga 100(n_\mathrm{H}/10^3~\mathrm{cm}^{-3})^{-1}$
Myr,
where $n_\mathrm{H}$ is the number density of hydrogen
nuclei in the cloud.}
We also examine the
relations between each of the extinction curve features (UV
slope, far-UV
curvature, and carbon bump strength) and $R_V$.
The correlation between UV slope and $R_V$,
which is the strongest among the
three correlations, is well reproduced by the tuned
model. For far-UV curvature and carbon bump strength,
the observational data are located between the
tuned model and the original model without tuning,
implying that the large scatters in the observational data can be
explained by the sensitive response to the coagulation
efficiency. The overall success of the tuned model
indicates that accretion and coagulation are promising
mechanisms of producing the variation of extinction curves in the
Milky Way, although we do not exclude possibilities of other
dust-processing mechanisms changing extinction curves.
\end{abstract}


\section{Introduction}

The evolution of interstellar dust is one of the most
important problems in clarifying the galaxy evolution.
The essential features of dust properties are
grain size distribution and grain materials.
Dust grains actually modify the spectral energy
distribution of galaxies by reprocessing stellar
radiation into the far-infrared regions in a way
dependent on the extinction and the emission cross-section
\citep[e.g.][]{desert90}, both of which are mainly
determined by the grain size distribution and the grain
materials. Not only radiative processes but
also interstellar chemical reactions are affected by
dust properties: especially,
formation of molecular hydrogen predominantly
occurs on dust grains if the interstellar medium
(ISM) is enriched with dust
\citep[e.g.][]{cazaux04,yamasawa11}. These effects of dust
grains underline the importance of
clarifying the dust enrichment and the
evolution of dust properties in galaxies.

The evolution of dust in a galaxy is governed
by various processes \citep[e.g.][]{asano13}.
Dust grains are supplied by
stellar sources such as supernovae (SNe) and
asymptotic giant branch (AGB) stars
\citep[e.g.][]{bianchi07,nozawa07,yasuda12}.
In the Milky Way ISM,
the time-scale of dust destruction by SN shocks is
$\mbox{a few}\times 10^8$~yr (\citealt*{jones96};
but see \citealt{jones11}), which is significantly
shorter than the time-scale
of dust supply from stellar sources ($\sim 1.5$ Gyr)
\citep{mckee89}.
Therefore, it has been argued that dust grains grow
in the ISM by the accretion of metals\footnote{We
call the elements composing
grains `metals'.}
(simply called accretion in this paper)
to explain the dust abundance in the ISM in various
types of galaxies
\citep{dwek98,zhukovska08,draine09,pipino11,valiante11,
inoue11,asano12}. The most efficient sites of grain
growth are molecular clouds,
where the typical number density of
hydrogen molecules is $\sim 10^3$ cm$^{-3}$
\citep{hirashita00}.

Although accretion is suggested to govern
the dust abundance, direct evidence of accretion is
still poor. Larger depletion of metal elements
in the cold clouds than in the warm medium
\citep{savage96} may indicate grain growth by
accretion in clouds.
\citet[hereafter VH10]{voshchinnikov10} show that the metals are
more depleted on to the dust as the optical extinction
curve becomes flatter. The depletion is an
indicator of how much fraction of the metals are
accreted on the
dust, while the flatness of extinction curve reflects
the size of dust grains. Thus, their result indicates
that the accretion of metals on to dust grains and
the growth of grain size occur at the same sites.

Dust grains also grow through coagulation; i.e.\
grain--grain sticking \citep[e.g.][]{chokshi93}.
A deficit of very small grains contributing to
the 60-$\micron$ emission is observed around
a typical density in molecular clouds
$\sim 10^3$ cm$^{-3}$, and is interpreted to be
a consequence of coagulation
\citep{stepnik03}. Coagulation efficiently occurs
in the dense medium like accretion. Thus,
it is reasonable to consider that accretion
and coagulation take place at the same time.

Although accretion and coagulation can simultaneously
occur, they may have different influences on
observational quantities. While accretion changes
both the total dust mass and the grain sizes,
coagulation only changes
the grain sizes, conserving the
total dust mass. In the above, we have mentioned
the observed relation
between depletion and extinction curve by VH10.
In interpreting
this relation, we should note that the depletion is
only affected by
accretion while the extinction curve is
influenced by both accretion and coagulation.
Therefore, there is a possibility of
disentangling the effects of accretion and coagulation
by interpreting the relation between
depletion and extinction curve.

The shape of extinction curve may be capable of
discriminating the two grain growth mechanisms,
accretion and coagulation
\citep*{cardelli89,odonnell97}.
Coagulation makes the grain sizes larger
and the extinction curve flatter, while accretion
increases the extinction itself.
\citet[][hereafter H12]{hirashita12} points out a
possibility that
accretion steepens the extinction curve at
ultraviolet (UV) and optical wavelengths.
%%This is explained as follows: accretion enhances the
%%abundance of the smallest grains, and since
%%the UV extinction is more sensitive to the
%%enhancement of small grains than the extinction at longer
%%wavelengths, the extinction curve is steepened by
%%accretion.
Thus, accretion and coagulation,
although these two processes occur simultaneously
in the dense ISM, can have different impacts on
the extinction curve.

The above arguments are mainly based on qualitative
observational evidence or purely theoretical arguments
on each of the two processes (accretion and coagulation).
Thus, in this paper, we solve the
evolution of extinction curve by accretion and
coagulation \textit{simultaneously} and compare with
some \textit{quantitative} indicators that characterize
major features of extinction curve. We also use
the metal depletion as an indicator of accretion, to
isolate the effect of accretion. As a consequence of
this study, we can address if the observed variation
of extinction curve is consistent with the grain growth
mechanisms and test the widely accepted hypothesis that
accretion and coagulation are really occurring in the ISM.
We concentrate on the Milky Way ISM,
for which detailed data are available for the
extinction curves.

This paper is organized as follows. We explain our
models used to calculate the effects of accretion and
coagulation on the extinction curve in
Section~\ref{sec:model}. We show the calculation
results and compare them with the observational data
of Milky Way extinction curves and depletion
in Section \ref{sec:results}. The models are also tested
by a large sample of Milky Way extinction curves
in Section \ref{sec:fitzpatrick}.
After discussing some implications of our results in
Section~\ref{sec:discussion},
we conclude in Section \ref{sec:conclusion}.

\section{Models}\label{sec:model}

We consider the time evolution of grain size distribution
by the following two processes:
the accretion of metals
and the growth by coagulation in an interstellar cloud.
These two processes are
simply called accretion and coagulation in
this paper. Our aim is to investigate if dust grains
processed by these processes can explain the
variation of extinction curves in various lines of sight.
We only treat grains
refractory enough to survive after the dispersal of
the cloud, and do not consider volatile grains
such as water ice.
We assume that the grains are spherical with
a constant material density $s$ dependent on
the grain species, so that the grain
mass $m$ and the grain radius $a$ are related as
\begin{eqnarray}
m=\frac{4}{3}\pi a^3s.\label{eq:mass}
\end{eqnarray}
We define the grain size distribution such that
$n(a,\, t)\,\mathrm{d}a$ is the number density of
grains whose radii are between $a$ and
$a+\mathrm{d}a$ at time $t$. In our numerical scheme,
we use the number density of grains with mass
between $m$ and $m+\mathrm{d}m$,
$\tilde{n}(m,\, t)$, which is related to
$n(a,\, t)$ by
$\tilde{n}(m,\, t)\,\mathrm{d}m=n(a,\, t)\,\mathrm{d}a$;
that is, $\tilde{n}=n/(4\pi a^2s)$.
For convenience, we also define the dust mass
density, $\rho_\mathrm{d}(t)$:
\begin{eqnarray}
\rho_\mathrm{d}(t)=\int_0^\infty\frac{4}{3}\pi a^3s\,
n(a,\, t)\,\mathrm{d}a.\label{eq:rho_d}
\end{eqnarray}

The time evolution of grain size distribution by
accretion and coagulation is calculated based on
the formulation by H12. We consider
silicate and carbonaceous dust as grain species.
To avoid the complexity in compound species, we
treat these two grain species separately.
{We neglect the effect of the Coulomb interaction
on the cross-section (i.e.\ the cross-section of a grain for
accretion and coagulation is simply estimated by the
geometric one).
For ionized particles,
the effect of electric polarization could raise the cross-section
especially for small grains \citep[e.g.][]{draine87}.
Therefore, neglecting the Coulomb interaction may lead to
an underestimate of grain growth by accretion. However,
the ionization degree
in dense clouds is extremely low \citep{yan04}, which means
that almost all the metal atoms colliding with the dust grains
are neutral. For
grain--grain collisions, the effect of charing is of minor
significance, since the kinetic energy of the grains is much
larger than the Coulomb energy at the grain surface.}
We briefly review the formulation by H12 in the
following. Accretion and coagulation, described separately
in the following subsections, are solved
simultaneously in the calculation.

\subsection{Accretion}

The equation for the time evolution of grain size
distribution by accretion is written as
(see H12 for the derivation)
\begin{eqnarray}
\frac{\partial\sigma}{\partial t}+\dot{\mu}
\frac{\partial\sigma}{\partial\mu}=\frac{1}{3}
\dot{\mu}\sigma,\label{eq:continuity_sigma}
\end{eqnarray}
where $\sigma (m,\, t)\equiv m\tilde{n}(m,\, t)$,
$\mu\equiv\ln m$, and
$\dot{\mu}\equiv\mathrm{d}\mu /\mathrm{d}t$.
The time evolution of the logarithmic grain mass
$\mu$ can be written as
\begin{eqnarray}
\dot{\mu}=\frac{3\xi_\mathrm{X}(t)}{\tau (m)},\label{eq:dmudt}
\end{eqnarray}
where $\tau (m)$ is the growth time-scale of grain radius
as a function of grain mass (given by equation \ref{eq:tau};
note that $m$ and $a$ are related by equation \ref{eq:mass}),
and $\xi_\mathrm{X}(t)=n_\mathrm{X}/n_\mathrm{X,tot}$
the gas-phase fraction of key element X (X = Si and C for
silicate and carbonaceous dust, respectively)
as a function of time
($n_\mathrm{X}$ is the number density of key species X in gas
phase as a function of time, and $n_\mathrm{X,tot}$
is the number density of element X in both gas and
dust phases). The growth time-scale is
evaluated as
\begin{eqnarray}
\tau (a)\equiv
\frac{a}{{\displaystyle
\frac{n_\mathrm{X,tot}\, m_\mathrm{X}S_\mathrm{acc}}
{f_\mathrm{X}s}}
\left({\displaystyle
\frac{k_\mathrm{B}T_\mathrm{gas}}{2\pi m_\mathrm{X}}}
\right)^{1/2}},\label{eq:tau}
\end{eqnarray}
where $m_\mathrm{X}$ is the atomic mass of X,
$S_\mathrm{acc}$ is the sticking probability for
accretion, $f_\mathrm{X}$
is the mass fraction of the key species in
dust, $k_\mathrm{B}$ is the Boltzmann
constant, and
$T_\mathrm{gas}$ is the gas temperature.
The evolution of $\xi_\mathrm{X}$ is calculated by
\begin{eqnarray}
\frac{\mathrm{d}\xi_\mathrm{X}}{\mathrm{d}t}=
\frac{-3f_\mathrm{X}\xi_\mathrm{X}(t)}{m_\mathrm{X}n_\mathrm{X,tot}}
\int_0^\infty\frac{\sigma (m,\, t)}{\tau (m)}\,\mathrm{d}m.
\label{eq:dxidt}
\end{eqnarray}
We solve equations (\ref{eq:continuity_sigma})--(\ref{eq:dxidt})
to obtain the time evolution of $\sigma (m,\, t)$, which is
translated into $n(a,\, t)$.

To clarify the time-scale of accretion, we give the
numerical estimates for $\tau$. With the values of
parameters given in Table \ref{tab:material},
$\tau (a)$ in equation (\ref{eq:tau})
can be estimated as
\begin{eqnarray}
\tau & = & 1.80\times 10^8\left(\frac{a}{0.1~\micron}
\right)\left(\frac{Z}{\mathrm{Z}_{\sun}}\right)^{-1}
\left(\frac{n_\mathrm{H}}{10^3~\mathrm{cm}^{-3}}
\right)^{-1}\nonumber\\
& & \times\left(\frac{T_\mathrm{gas}}{10~\mathrm{K}}
\right)^{-1/2}\left(\frac{S_\mathrm{acc}}{0.3}\right)^{-1}~
\mathrm{yr}
\label{eq:tau_sil}
\end{eqnarray}
for silicate, and
\begin{eqnarray}
\tau & = & 0.993\times 10^8\left(\frac{a}{0.1~\micron}
\right)\left(\frac{Z}{\mathrm{Z}_{\sun}}\right)^{-1}
\left(\frac{n_\mathrm{H}}{10^3~\mathrm{cm}^{-3}}
\right)^{-1}\nonumber\\
& & \times\left(\frac{T_\mathrm{gas}}{10~\mathrm{K}}
\right)^{-1/2}\left(\frac{S_\mathrm{acc}}{0.3}\right)^{-1}~
\mathrm{yr}
\label{eq:tau_gra}
\end{eqnarray}
for carbonaceous dust
{($n_\mathrm{H}$ is the number density of hydrogen
nuclei)}. We adopt the same values as
those in H12 for the following quantities:
$Z=\mathrm{Z}_{\sun}$, $n_\mathrm{H}=10^3$ cm$^{-3}$,
$T_\mathrm{gas}=10$~K, and $S_\mathrm{acc}=0.3$.
Note that the same result is obtained with the
same value of $Zn_\mathrm{H}St$; in particular,
the density may vary in a wide range, so that
it is worth noting that the time-scales of
accretion and coagulation both scale with
$\propto n_\mathrm{H}^{-1}$
{(see also Section \ref{subsec:lifetime})}.

\begin{table}
\centering
\begin{minipage}{80mm}
\caption{Adopted quantities.}
\label{tab:material}
    \begin{tabular}{lccccc}
     \hline
     Species & X & $f_\mathrm{X}\,^{a}$ &
     $m_\mathrm{X}\,^\mathrm{b}$
     & (X/H)$_{\sun}\,^\mathrm{c}$ &
     $s\,^\mathrm{c}$\\
      & & & amu & & (g cm$^{-3}$) \\
     \hline
     Silicate & Si & 0.163 & 28.1 & $3.3\times 10^{-5}$ & 3.5 \\
     Graphite & C  & 1     & 12 & $3.63\times 10^{-4}$ & 2.24 \\
     \hline
    \end{tabular}
    
    \medskip

%%\textit{Note.} All the quantities follow HK11.\\
$^{a}$For silicate, we assume a composition of
MgFeSiO$_4$ \citep{weingartner01}.\\
$^{b}$The atomic masses and the abundances
are taken from \citet{cox00}.\\
$^{c}$We adopt the same values as those used by
\citet{weingartner01} for the models. Although the
solar abundance varies among various measurements,
such a variation does not affect our conclusions.
\end{minipage}
\end{table}

\subsection{Coagulation}\label{subsec:coag}

We solve a discretized coagulation equation used in
\citet{hirashita09}. We consider thermal (Brownian)
motion and turbulent motion as a function of grain mass
(or radius, which is related to the mass by
equation \ref{eq:mass}) (see H12 for details). The
size-dependent velocity dispersion
of grains is denoted as $v(a)$.
The coagulation is assumed to occur if the
relative velocity is below the coagulation
threshold given by \citet{hirashita09}
(see \citealt{chokshi93,dominik97,yan04}) unless
otherwise stated.
%%For a grain colliding
%%with $a\sim 0.01~\micron$ grains, coagulation
%%occurs if the grain has a size smaller than
%%$\mbox{a few}\times 0.01~\micron$.
We modify the treatment of relative velocities
following \citet{hirashita13}: in
considering a collision between grains with
sizes $a_1$ and $a_2$, we estimate the
relative velocity, $v_{12}$ by
\begin{eqnarray}
v_{12}=\sqrt{v(a_1)^2+v(a_2)^2-2v(a_1)v(a_2)\cos\theta\,},
\end{eqnarray}
where $\theta$ is the angle between the two grain
velocities, and $\cos\theta$ is randomly chosen
between $-1$ and 1 in each time-step. We also
assume that the sticking coefficient $S_\mathrm{coag}$
is 1 unless otherwise stated, also examining
models (the tuned model described later) in which
$S_\mathrm{coag}$ is less than 1.

\subsection{Initial conditions}\label{subsec:initial}

We adopt an initial grain size distribution that
fits the mean Milky Way extinction curve.
\citet{weingartner01} assume the following functional
form for the grain size distributions of silicate and
carbonaceous dust:
\begin{eqnarray}
\lefteqn{
n(a)/n_\mathrm{H}=\frac{C_i}{a}\left(\frac{a}{a_{t,i}}
\right)^{\alpha_i}F(a;\,\beta_i,\, a_{t,i})
}\nonumber\\
& \times & \left\{
\begin{array}{ll}
1 & (3.5~\mathrm{\AA} <a<a_{t,i});\\
\exp\left\{ -[(a-a_{t,i})/a_{c,i}]^3\right\} & (a>a_{t,i});
\end{array}
\right.\label{eq:initial}
\end{eqnarray}
where $i$ specifies the grain species
($i=s$ for silicate and $i=g$ for carbonaceous dust,
which is assumed to be graphite) and the term
\begin{eqnarray}
F(a;\,\beta_i,\, a_{t,i})\equiv\left\{
\begin{array}{ll}
1+\beta a/a_{t,i} & (\beta\geq 0);\\
(1-\beta a/a_{t,i})^{-1} & (\beta <0)
\end{array}
\right.
\end{eqnarray}
provides curvature.
We omit the log-normal term for
small carbonaceous dust (i.e.\ $b_\mathrm{C}=0$ in
\citealt{weingartner01}), which is dominated by
polycyclic aromatic hydrocarbons (PAHs), since it is not
clear how PAHs contribute to the accretion and
coagulation to form macroscopic grains.
{Note that \citet{weingartner01} favour $b_\mathrm{C}>0$
based on the observed strong PAH emission from the
diffuse ISM \citep{li01}. If PAHs contribute to form
macroscopic grains through
accretion and coagulation, our estimate gives a conservative
estimate for the effect of accretion. However, after
coagulation becomes dominant at $\ga 100$ Myr,
the existence of PAHs does not affect the results
as long as PAHs have a negligible contribution to the total
dust mass, which is the case in \citet{weingartner01}'s
grain size distributions.}
%%If $b_\mathrm{C}>0$, accretion occurs more efficiently,
%%but as shown later, too efficient accretion enhance
%%the carbon bump at 0.22 $\micron$ too much.
The size distribution has five parameters
$(C_i,\, a_{t,i},\, a_{c,i},\,\alpha_i,\,\beta_i)$
for each grain species. We adopt the solution for
$b_\mathrm{C}=0$ and $R_V=3.1$
(see equation \ref{eq:rv} for the definition of $R_V$)
in \citet{weingartner01}
to fix these
parameters. The adopted values are listed in
Table \ref{tab:initial}.

\begin{table*}
\centering
\begin{minipage}{100mm}
\caption{Parameter values for the initial grain size
distributions.}
\label{tab:initial}
\begin{tabular}{lccccc}
\hline
Species & $C_i$ & $a_{t,i}$ & $a_{c,i}$ &
$\alpha_i$ & $\beta_i$ \\
 & & ($\micron$) & ($\micron$) & & \\
\hline
Silicate  & $1.02\times 10^{-12}$ & 0.172 & 0.1 & $-1.48$ &
$-9.34$ \\
Carbonaceous dust & $9.94\times 10^{-11}$ & 0.00745 & 0.606 &
$-2.25$ & $-0.0648$ \\
\hline
\end{tabular}
%%
%%\medskip
%%
%%$^\mathrm{a}$ 3 $\sigma$ upper limits with
%%$\Delta V=40$ km s$^{-1}$.
\end{minipage}
\end{table*}

The initial value of $\xi_\mathrm{X}$ is given
for each dust species as follows. The dust mass density
at $t=0$, $\rho_{\mathrm{d}}(0)$, is given
for silicate and carbonaceous dust separately
by equation~(\ref{eq:rho_d}) with the
initial grain size distribution
(equation \ref{eq:initial}). As derived in H12,
$\rho_\mathrm{d}(0)$ is related to the initial
condition for $\xi_\mathrm{X}$ as
\begin{eqnarray}
\rho_\mathrm{d}(0)=
\frac{m_\mathrm{X}}{f_\mathrm{X}}[1-\xi_\mathrm{X}(0)]
\left(\frac{Z}{\mathrm{Z}_{\sun}}\right)
\left(\frac{\mathrm{X}}{\mathrm{H}}\right)_{\sun}
n_\mathrm{H},\label{eq:rho_d0}
\end{eqnarray}
where $Z$ is the metallicity,
(we apply $Z=\mathrm{Z}_{\sun}$ throughout this
paper), and
$(\mathrm{X/H})_{\sun}$ is the solar abundance
(the ratio of the number of X nuclei to that of hydrogen
nuclei at the solar metallicity).
With the solar abundances of Si and C
(Table \ref{tab:material}), we
obtain $\xi_\mathrm{C}(0)=0.198$ for carbonaceous dust, while
$\xi_\mathrm{Si}(0)$ is negative for silicate. This means that
too much silicon is used to explain the Milky Way
extinction curve by \citet{weingartner01}.
As they mentioned, since there is still an
uncertainty in the interstellar elemental abundances,
we do not try to adjust the silicon and carbon
abundances. In fact, for the purpose of investigating
the effects of accretion and coagulation on the extinction
curve, fine-tuning of elemental abundance is not
essential. Thus, we assume $\xi_\mathrm{Si}(0)=0.1$
as a typical depletion of silicon
in the ISM (VH10), while
we adopt
$\xi_\mathrm{C}(0)=0.198$ as expected from the
interstellar carbon abundance.

\subsection{Calculation of extinction curves}
\label{subsec:ext}

Extinction curves are calculated by adopting the
method described by \citet{weingartner01}.
The extinction at wavelength $\lambda$ in units
of magnitude ($A_\lambda$) normalized to
the column density of hydrogen
nuclei ($N_\mathrm{H}$) is
written as
\begin{eqnarray}
\frac{A_\lambda (t)}{N_\mathrm{H}}=
\frac{2.5\log\mathrm{e}}{n_\mathrm{H}}
\int_0^\infty\mathrm{d}a\, n(a,\, t)
\pi a^2Q_\mathrm{ext}(a,\,\lambda ),\label{eq:A_N}
\end{eqnarray}
where $Q_\mathrm{ext}(a,\,\lambda )$ is the
extinction efficiency factor, which is evaluated
by using the Mie theory \citep{bohren83}. For
the details of optical constants for silicate and
carbonaceous dust, see \citet{weingartner01}.
The colour excess at wavelength $\lambda$ is defined as
\begin{eqnarray}
E(\lambda -V)\equiv A_\lambda-A_V,
\end{eqnarray}
where the wavelength at the $V$ band is 0.55 $\micron$.
The steepness of extinction curve is often quantified
by $R_V$:
\begin{eqnarray}
R_V\equiv\frac{A_V}{E(B-V)},\label{eq:rv}
\end{eqnarray}
where the wavelength at the $B$ band is 0.44 $\micron$.
Since a flat extinction curve decreases
the difference between $A_V$ and $A_B$ [i.e.\ $E(B-V)$],
$R_V$ increases
as the extinction curve becomes flatter.

To extract some features in UV extinction curves,
we apply a parametric fit to the calculated extinction
curves according to \citet{fitzpatrick07}. They adopt
the following form
for $\lambda\leq 2700$~\AA:
\begin{eqnarray}
\lefteqn{E(\lambda -V)/E(V-B)}\nonumber\\
& & \hspace{-5mm}= \left\{
\begin{array}{ll}
c_1+c_2x+c_3D(x,\, x_0,\,\gamma ) & (x\leq c_5),\\
c_1+c_2x+c_3D(x,\, x_0,\,\gamma )+c_4(x-c_5)^2 &
(x>c_5),
\end{array}
\right.\nonumber\\
& &
\end{eqnarray}
where $x\equiv\lambda^{-1}~(\micron^{-1})$, and
\begin{eqnarray}
D(x,\, x_0,\,\gamma )=
\frac{x^2}{(x^2-x_0^2)^2+x^2\gamma^2}
\end{eqnarray}
is introduced to express the 2175-\AA\ bump
(carbon bump).
There are seven parameters,
$(c_1,\, ...,\, c_5,\, x_0,\,\gamma )$.
To stabilize the fitting, we fix
$c_5=6.097$, $x_0=4.592~\micron^{-1}$, and
$\gamma =0.922~\micron^{-1}$ by adopting the
values fitting to the mean extinction curve
in the Milky Way \citep{fitzpatrick07}. The ranges
of these parameters are narrow compared with
the other parameters.
We apply a least-square fitting to the
theoretically predicted
extinction curves to derive
($c_1$, ..., $c_4$).

It is observationally suggested that there is a link
between UV extinction and $R_V$
\citep{cardelli89} although the scatter is large
\citep{fitzpatrick07}. The carbon
bump strength has also a weak correlation with
$R_V$ in such a way that the bump is smaller for
a flatter extinction curve (larger $R_V$)
\citep{cardelli89,fitzpatrick07}. We examine if
these relations are consistent with
accretion and coagulation. To estimate the bump
strength, we introduce the following two quantities:
$A_\mathrm{bump}\equiv \pi c_3/(2\gamma )$ and
$E_\mathrm{bump}\equiv c_3/\gamma^2$
(note that $A_\mathrm{bump}$ and
$E_\mathrm{bump}$ are simply determined by
$c_3$ after fixing $\gamma$). If an extinction curve is
steep in the optical it tends to be steep also in
the UV. This is reflected in the correlation between
$R_V$ and $c_2$ ($c_2$ is the slope of
UV extinction curve at both sides of
the carbon bump). Following \citet{fitzpatrick07},
we also define a far-UV (FUV) curvature as
$\Delta 1250\equiv c_4(8.0-c_5)^2$.

\subsection{Data}\label{subsec:data}

We test the change of grain size distribution by
accretion and coagulation through the comparison
of our models with observational extinction and
depletion data compiled by VH10 for stars located
in the Scorpius-Ophiuchus region, since the data
covers a wide range in $R_V$. There are 7 stars
(HD 143275, HD 144217, HD 144470, HD 147165,
HD 147888, HD 147933, and HD 148184): for these
stars, Mg or Si abundance in the line of sight is
measured. The depletion derived from the metal
abundance measurements is crucial to separate
the effect of accretion.
Among the 7 stars, extinction curves are available
for 4 stars (HD 144470, HD 147165,
HD 147888, and HD 147933) \citep{voshchinnikov12}.

We also use a larger sample compiled by
\citet{fitzpatrick07}, although they do not have
metal abundance data. They compiled UV-to-near-infrared
extinction data for
Galactic 328 stars. In particular, some important
parameters that characterize the extinction curve
($A_\mathrm{bump}$, $E_\mathrm{bump}$, $c_2$, and
$\Delta 1250$) are available, so that we can
check if the change of extinction curve by
accretion and coagulation is consistent with the
observed trend or range of these parameters for
various $R_V$.

\section{Results}\label{sec:results}

\subsection{Extinction curves}\label{subsec:extinc}

\subsubsection{General behaviour of theoretical
predictions}

\begin{figure*}
\includegraphics[width=0.45\textwidth]{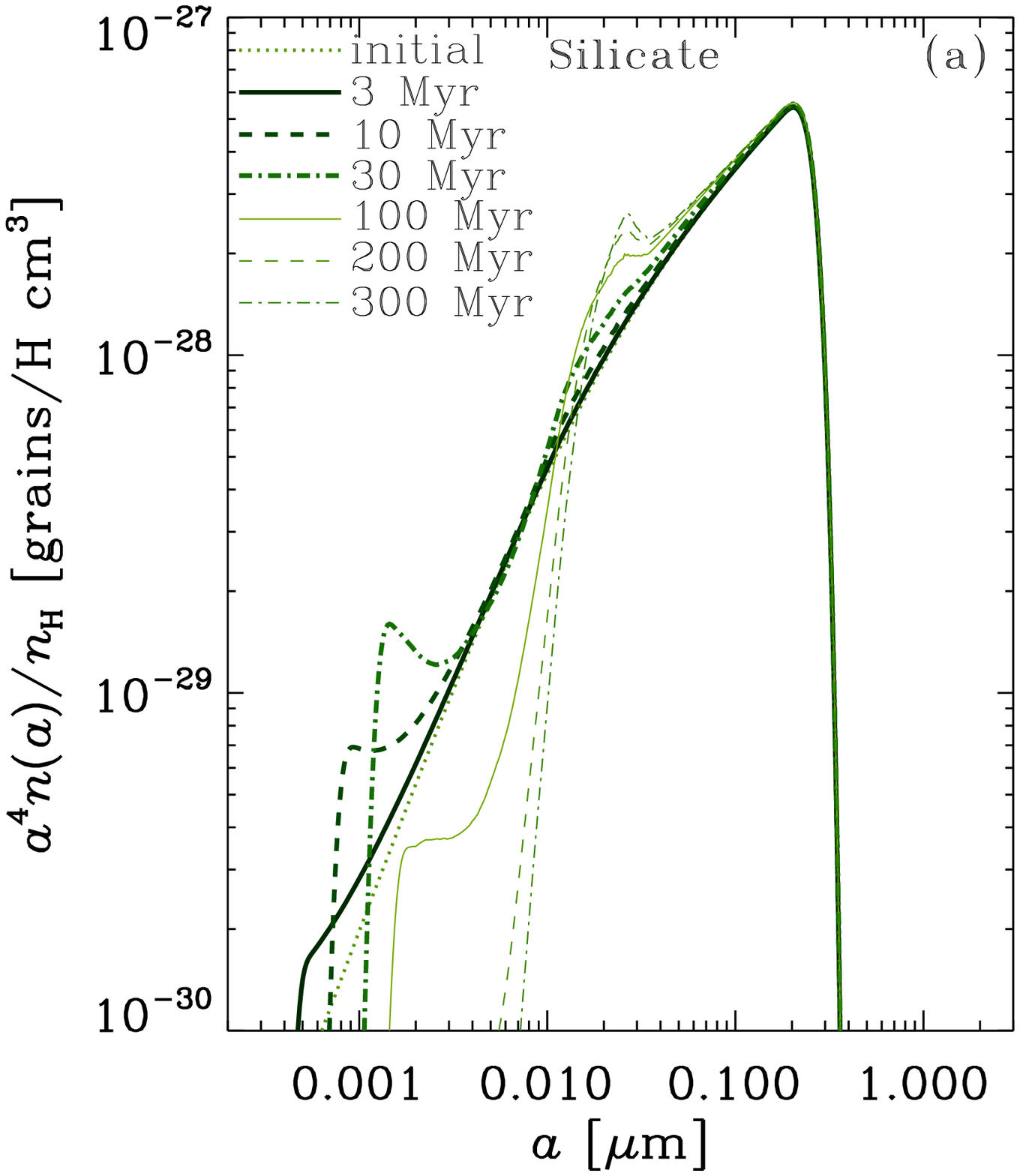}
\includegraphics[width=0.45\textwidth]{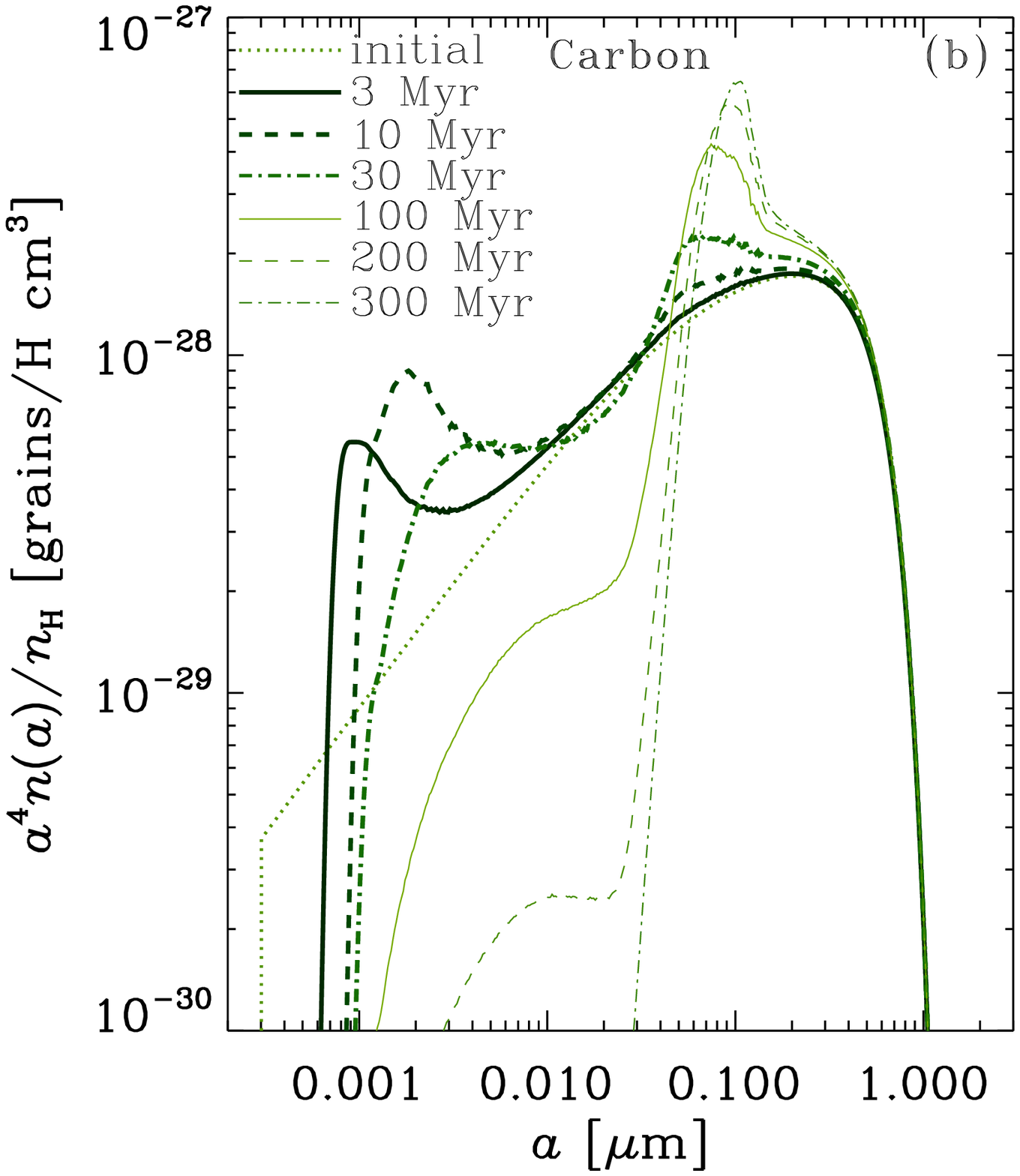}
\caption{Evolution of grain size distribution for
(a) silicate and (b) carbonaceous dust.
The grain size distribution is normalized to the
hydrogen number density and multiplied by $a^4$ to
show the mass distribution per logarithmic grain
radius bin.
The thick solid, thick dashed, thick dot-dashed,
thin solid, thin dashed, and thin dot-dashed lines
show the grain
size distributions at $t=3$, 10, 30, 100, 200 and 300 Myr,
respectively. The initial condition is shown by the
dotted line.
\label{fig:size}}
\end{figure*}

We show the evolution of grain size distribution in
Fig.\ \ref{fig:size} for silicate and carbonaceous dust.
To show the mass distribution per logarithmic size,
we multiply ${a}^4$ to ${n}$. The details about
how accretion and coagulation contribute
to the grain size distribution have already been
discussed in H12. We briefly overview the behaviour.

Since the growth rate of grain radius by accretion is
independent of $a$ (H12), the
impact of grain growth is significant at small
grain sizes. Moreover, almost all gas-phase metals
accrete selectively on to small grains
because the contribution to the grain surface
is the largest at the smallest sizes
\citep[see also][]{weingartner99}. Since the
accretion time-scale for the smallest grains
($a\sim 0.001~\micron$) is short, the effect of accretion
appears at first, enhancing the grain abundance at
the smallest sizes. Accretion stops after the gas-phase
elements composing the dust are used up
($\sim$ 10--30 Myr).
After accretion stops, the evolution of grain size
distribution is driven by coagulation.
Coagulation occurs in a bottom-up manner, because
small grains dominate the total grain surface. Moreover,
the grains at the largest sizes are
intact because they have velocities larger than
coagulation threshold.

\begin{figure*}
\includegraphics[width=0.45\textwidth]{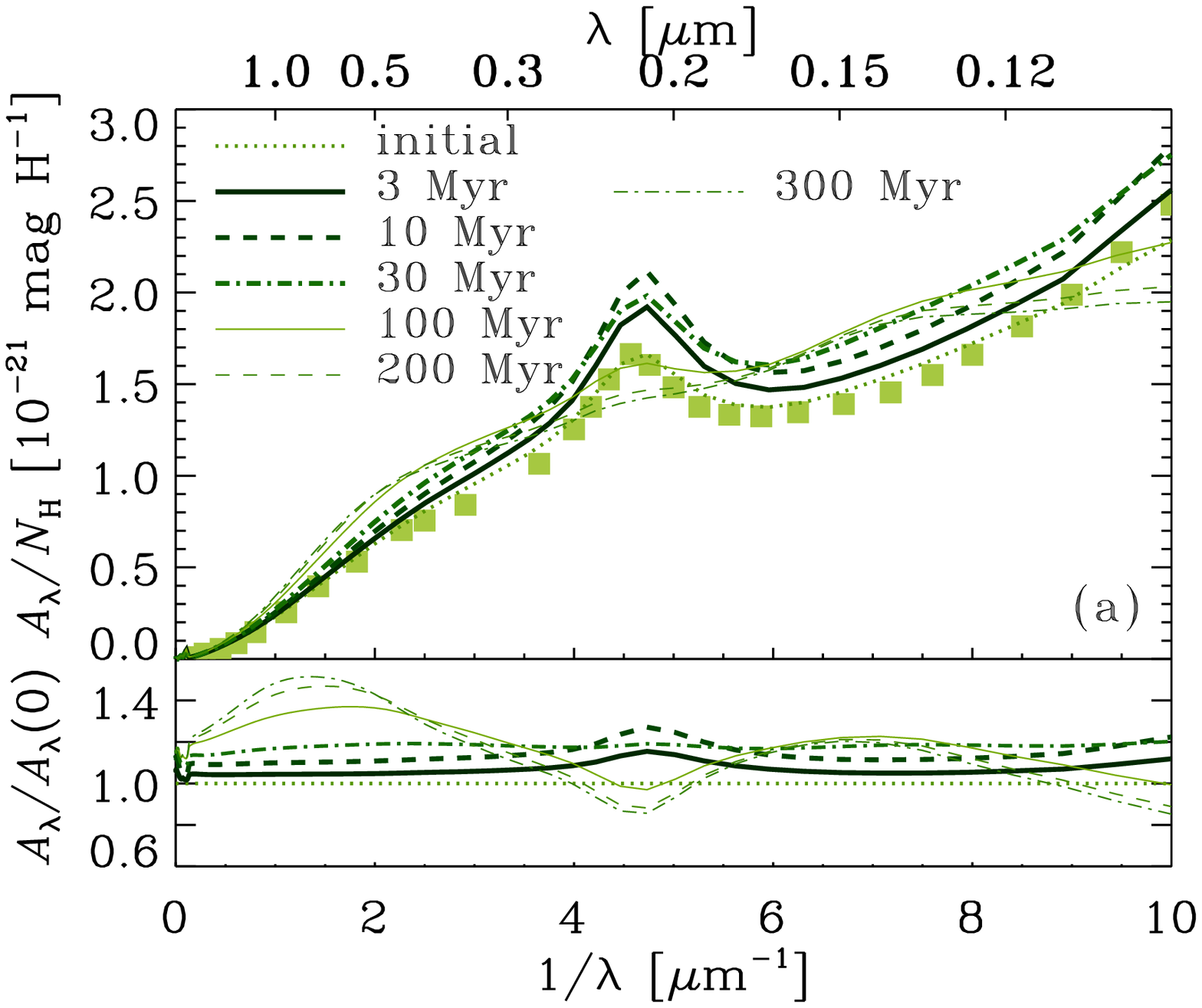}
\includegraphics[width=0.45\textwidth]{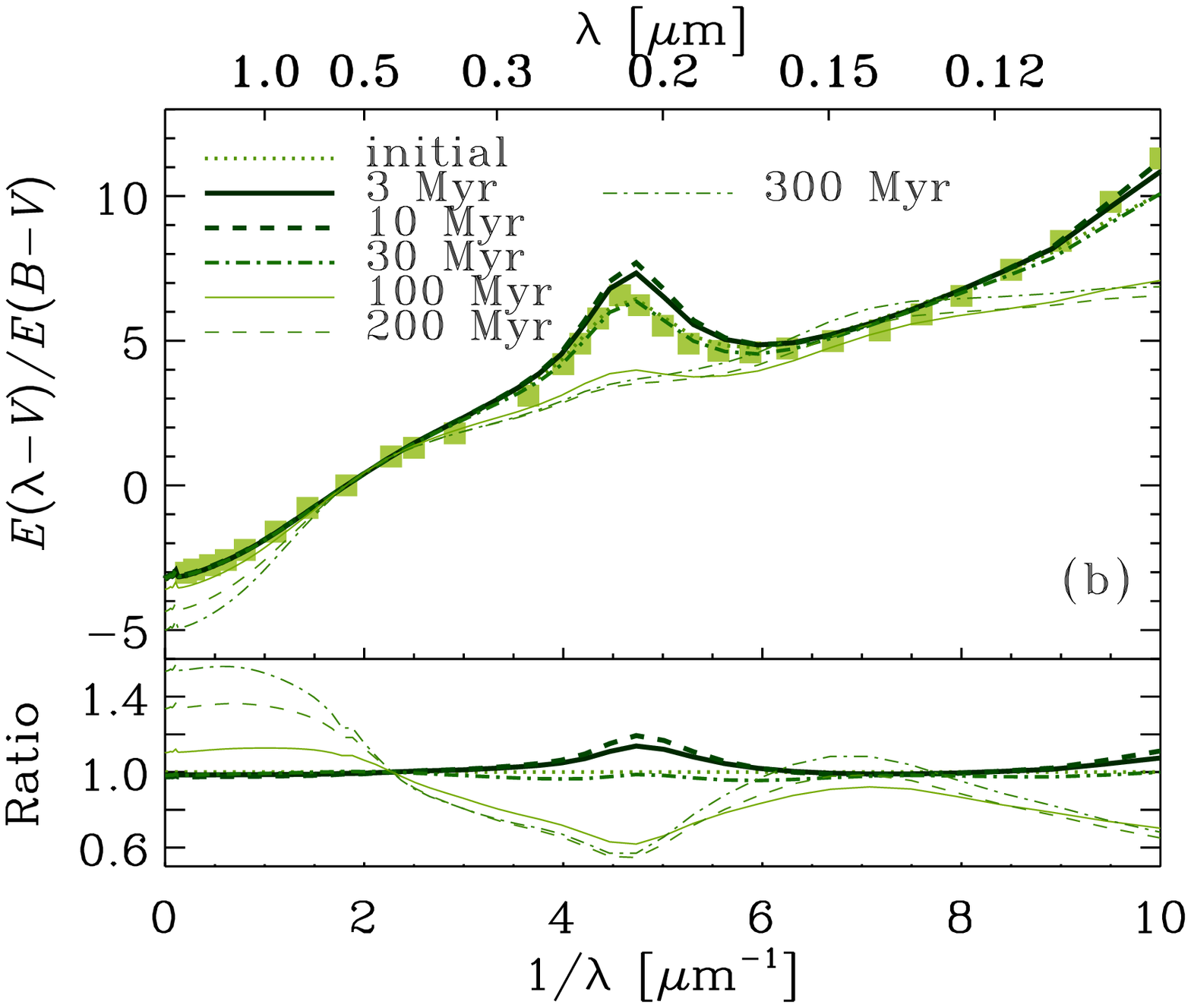}
\caption{
Evolution of extinction curves. We show
(a) $A_\lambda/N_\mathrm{H}$, and (b) $E(\lambda -V)/E(B-V)$.
The dotted line is the initial extinction curve
before accretion and coagulation.
The thick solid, thick dashed, thick dot-dashed,
thin solid, thin dashed, and thin dot-dashed lines show the grain
size distributions at $t=3$, 10, 30, 100, 200, and 300 Myr,
respectively.
The filled squares represent the mean Milky Way extinction data
taken from \citet{pei92} as a reference.
In each panel, the lower window shows the extinction divided
by the initial extinction (the correspondence between the line
types and the models is the same as above).
The values of $R_V$ are 3.26, 3.21, 3.17, 3.20, 3.59, 4.36,
and 5.00 at $t=0$, 3, 10, 30, 100, 200, and 300 Myr, respectively.
\label{fig:ext}}
\end{figure*}

Based on the grain size distributions shown above,
we calculate the evolution of extinction curve.
In Fig.\ \ref{fig:ext}a, we show the extinction curves
($A_\lambda /N_\mathrm{H}$, calculated by
equation \ref{eq:A_N}) for various $t$.
To show that the initial condition reproduces the mean
extinction curve, we also plot the averaged extinction
curve taken from \citet{pei92} (the shape is almost
identical to \citet{fitzpatrick07}'s mean extinction curve).
We also show in Fig.\ \ref{fig:ext}b the normalized colour
excess, $E(\lambda -V)/E(B-V)$. In the bottom panels of those
two figures, we present all the curves divided by
the initial values to indicate how the steepness of
extinction curve changes as a function of time.

In Fig.\ \ref{fig:ext}a, we observe that the extinction
increases at all wavelengths at $t\la 10$ Myr because
accretion increases the dust abundance.
The extinction normalized to the value at $t=0$
[$A_\lambda /A_\lambda (0)$] increases toward
shorter wavelengths at $t\la 10$ Myr, indicating that
accretion steepens the UV extinction curve.
This is because the UV extinction responds
more sensitively to the increase of the grains
at $a\sim 0.001$--0.01 $\micron$ by accretion than
the optical--near-infrared extinction (see also H12).
After that,
coagulation makes the grains larger (Fig.\ \ref{fig:size}),
decreasing the UV extinction and flattening the
extinction curve. The optical--near-infrared extinction
continues to increase
even after $\sim 10$ Myr because the increase of
grain sizes by coagulation increases the grain
opacity in the optical and near-infrared.
The carbon (2175-\AA) bump, which is produced by small
($a\la 300$~\AA; \citealt{draine84})
carbonaceous grains, become less
prominent as coagulation proceeds, because
small carbonaceous grains responsible for the
bump are depleted after coagulation.

The normalized colour excesses shown in
Fig.\ \ref{fig:ext}b naturally follow the same
behaviour as in Fig.\ \ref{fig:ext}a. From
the bottom panel
in Fig.\ \ref{fig:ext}b,
we observe that the ratio is $>1$ at
wavelengths shorter than 0.44 $\micron$ while
accretion dominates ($t\la 10$ Myr); and that
it is $<1$ after coagulation takes
place significantly. This behaviour confirms that accretion
first makes the extinction curve steep and coagulation
then modifies it in the opposite way (see also H12).
The increase of $R_V$ by coagulation is also reflected
in the increase of the absolute value of the intercept,
$-\lim_{1/\lambda\to +0}E(\lambda -V)/E(B-V)
=R_V$.

\subsubsection{Comparison with observed extinction curves}

\begin{figure*}
\includegraphics[width=0.45\textwidth]{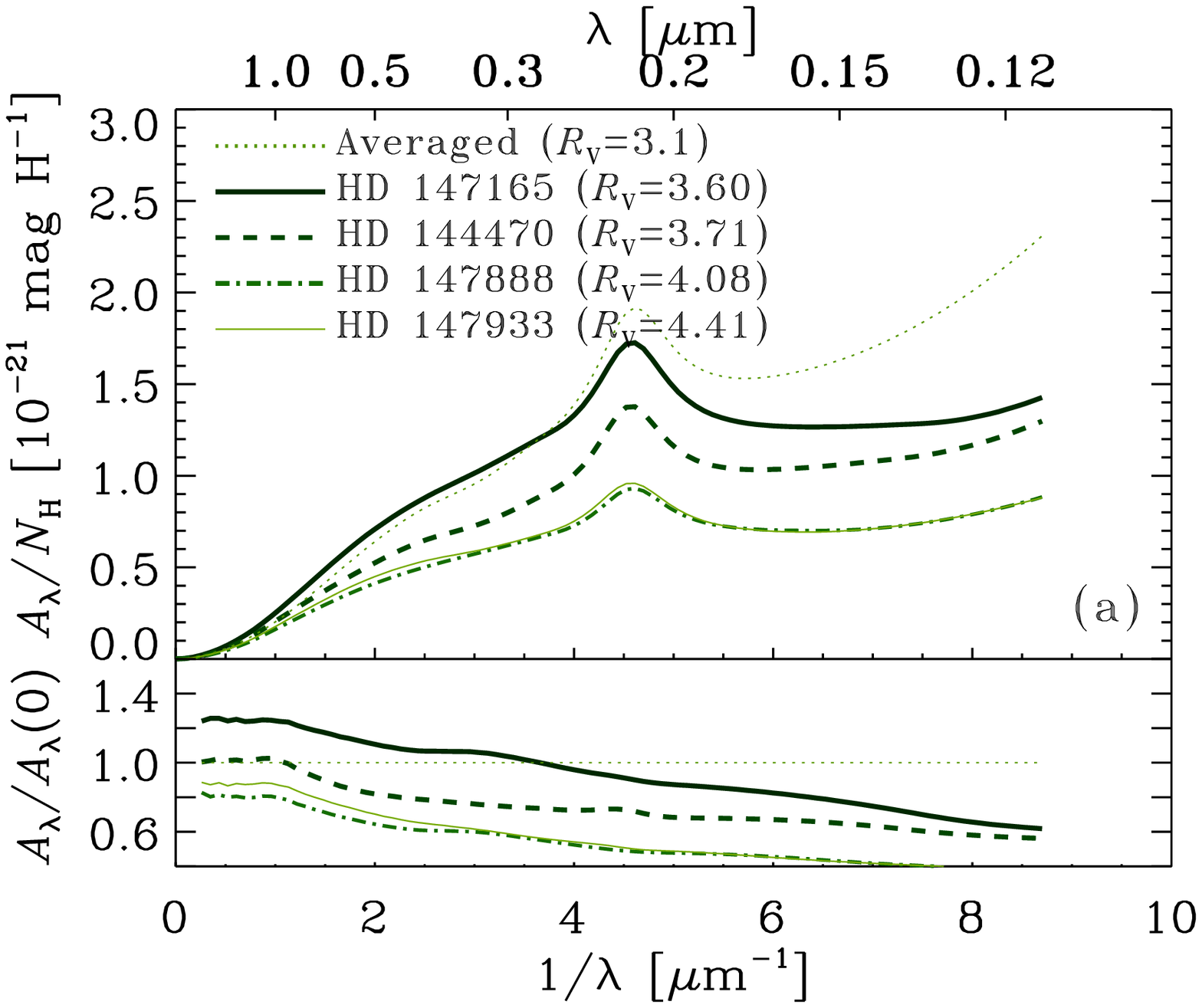}
\includegraphics[width=0.45\textwidth]{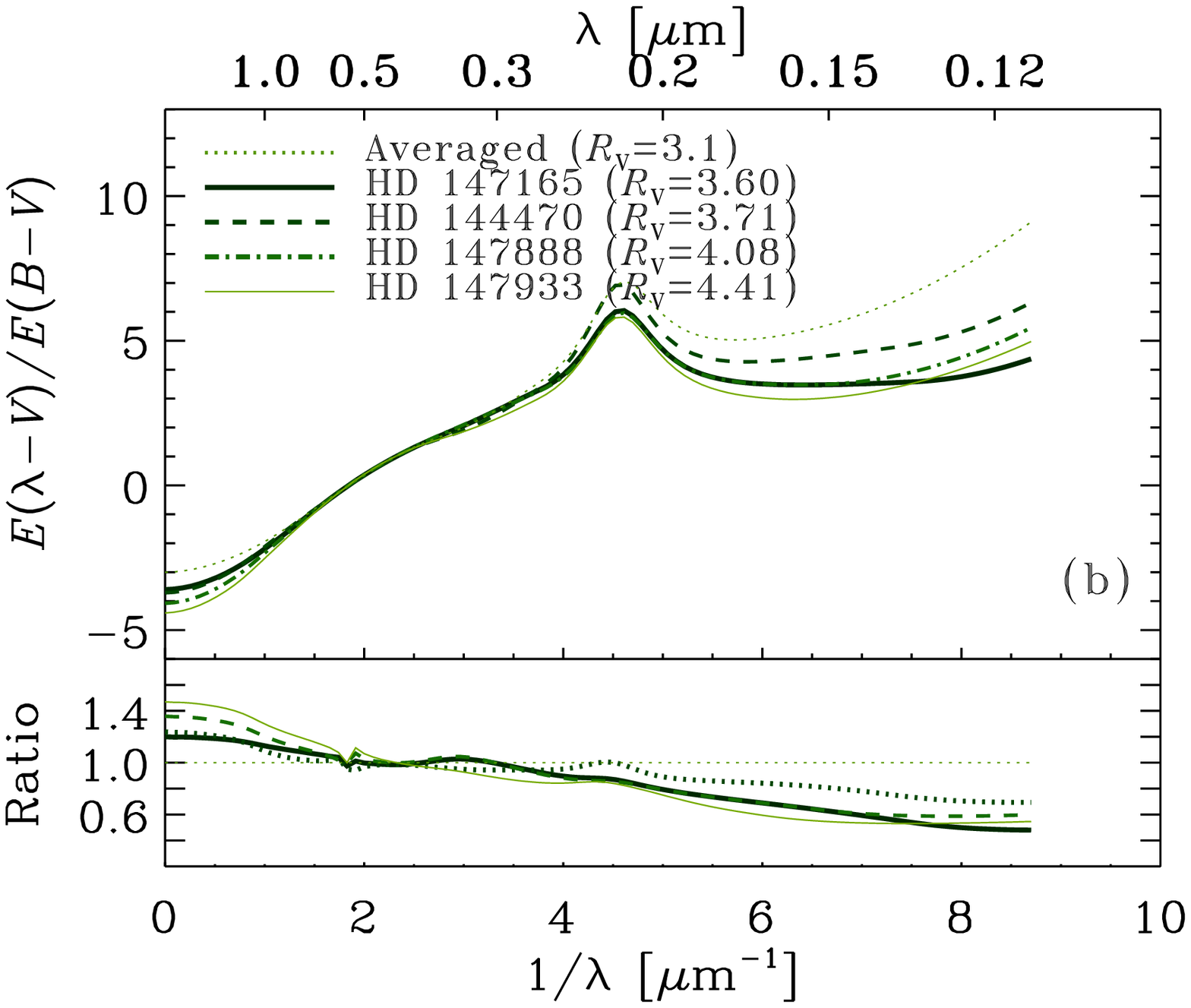}
\caption{
Observed extinction curves for our sample. We show
(a) extinction curves in units of magnitude
per hydrogen, $A_\lambda /N_\mathrm{H}$, and (b)
normalized colour excess, $E(\lambda -V)/E(B-V)$.
The thick solid, dashed, dot-dashed, and
thin solid lines show the extinction curves for
HD 147165, HD 144470, HD 147888, and HD 147933,
respectively. The dotted line represents
the averaged Milky Way curve for reference.
The $R_V$ values are also indicated.
In each panel, the lower window shows the extinction divided
by the averaged extinction (the correspondence between the line
types and the objects is the same as above).
\label{fig:ext_obs}}
\end{figure*}

For our sample (Section \ref{subsec:data}), extinction
curves are available for four stars. These extinction curves
are shown in Fig.\ \ref{fig:ext_obs}. Even for large $R_V$,
the carbon bump around 0.22 $\micron$ is clear and
the UV extinction rises with a positive curvature.
However, our theoretical predictions shown
in Fig.\ \ref{fig:ext} indicate that at large $R_V$
(i.e.\ after significant coagulation at $>100$ Myr),
the carbon bump disappears. This implies that coagulation
is in reality not so efficient as assumed in the model
for carbonaceous dust. There is another discrepancy:
the observed $A_V/N_\mathrm{H}$ tends to decrease
significantly as $R_V$ increases,
while it does not decrease significantly after significant
coagulation at $>100$ Myr, except at the carbon bump. This is
because silicate stops to coagulate at $a\sim 0.03~\micron$,
which is still too small to affect the UV opacity.

In summary, the observational data indicates
(i) that carbonaceous dust should be more inefficient in
coagulation than assumed, and (ii) that silicate should grow
beyond $a\sim 0.03~\micron$ (the growth is limited
by the coagulation threshold velocity).
Therefore, we also try models with
$S_\mathrm{coag}<1$ for carbonaceous dust and
no coagulation threshold for silicate dust as
a tuned model described in the next subsection.

\subsection{Tuned model}

\begin{figure*}
\includegraphics[width=0.45\textwidth]{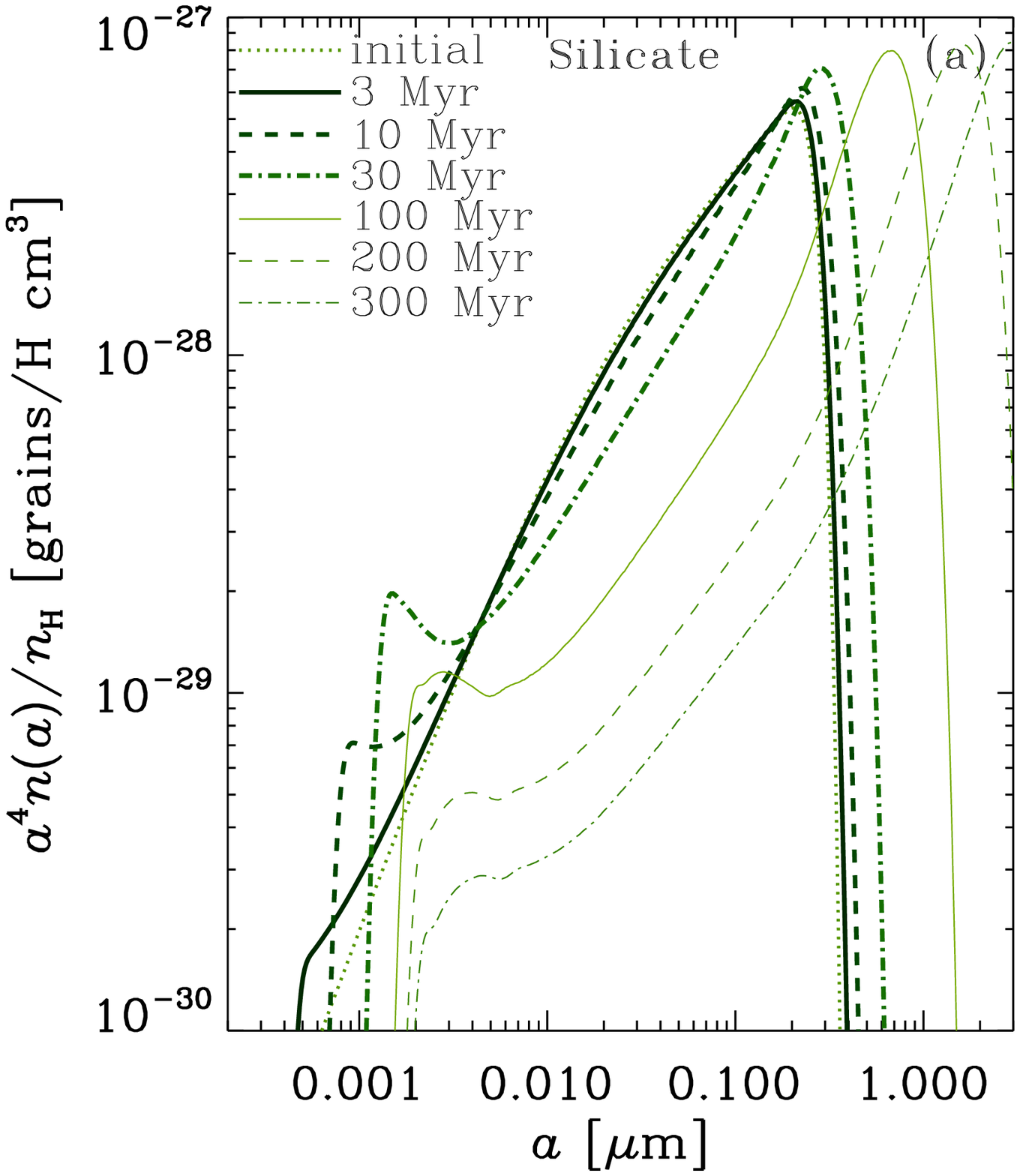}
\includegraphics[width=0.45\textwidth]{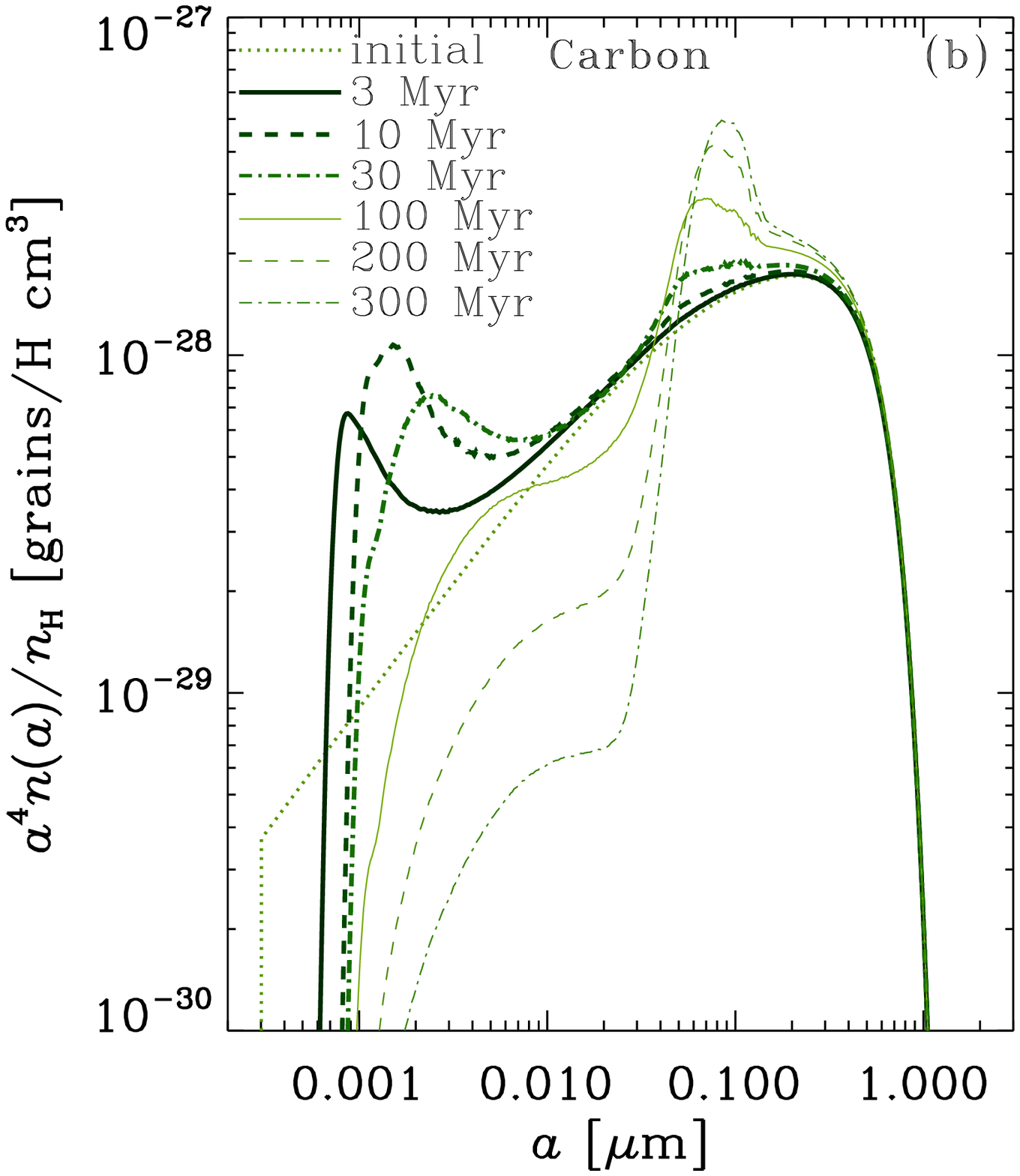}
\caption{
Same as Fig.\ \ref{fig:size} but for the tuned model.
\label{fig:size_tuned}}
\end{figure*}

\begin{figure*}
\includegraphics[width=0.45\textwidth]{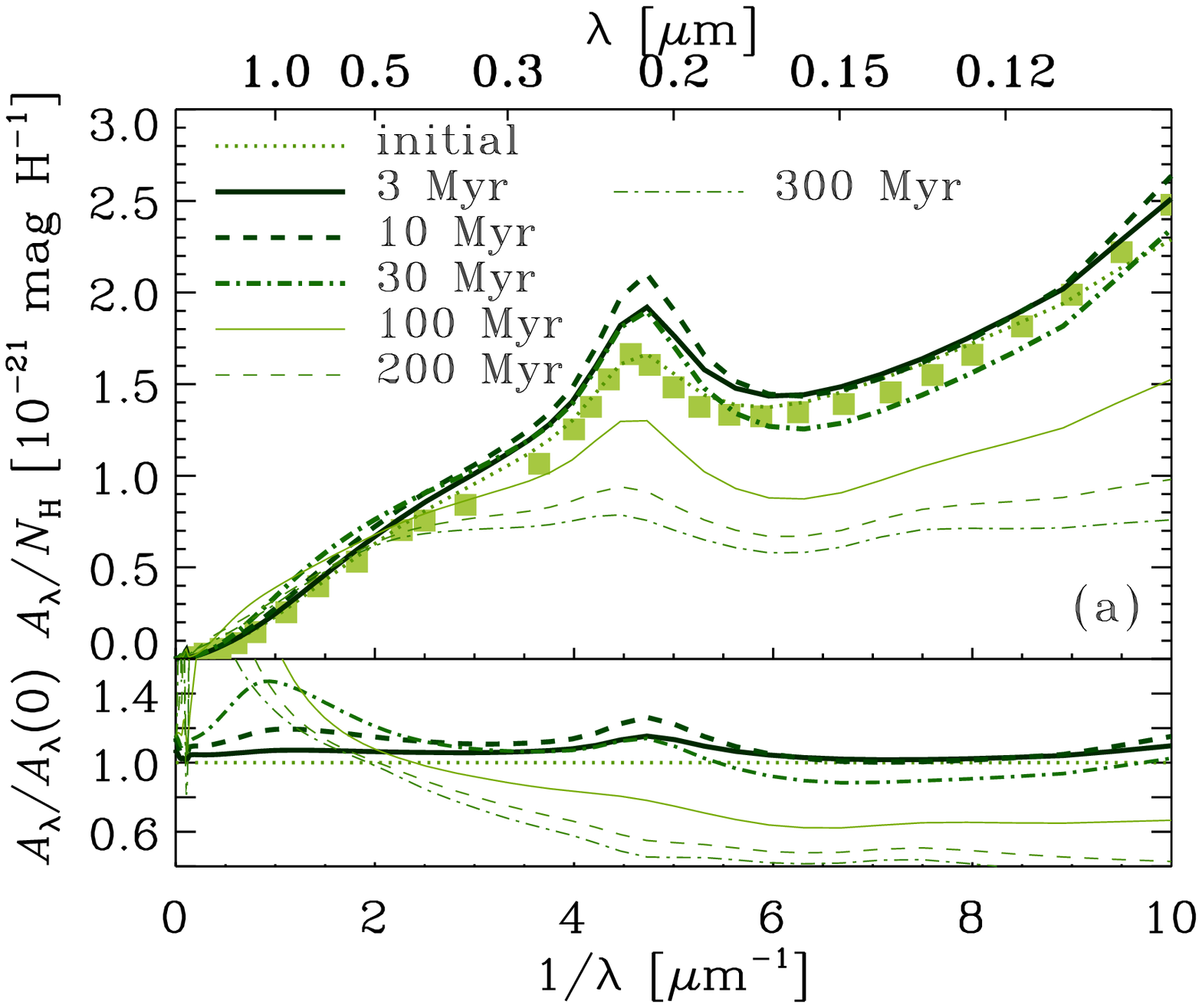}
\includegraphics[width=0.45\textwidth]{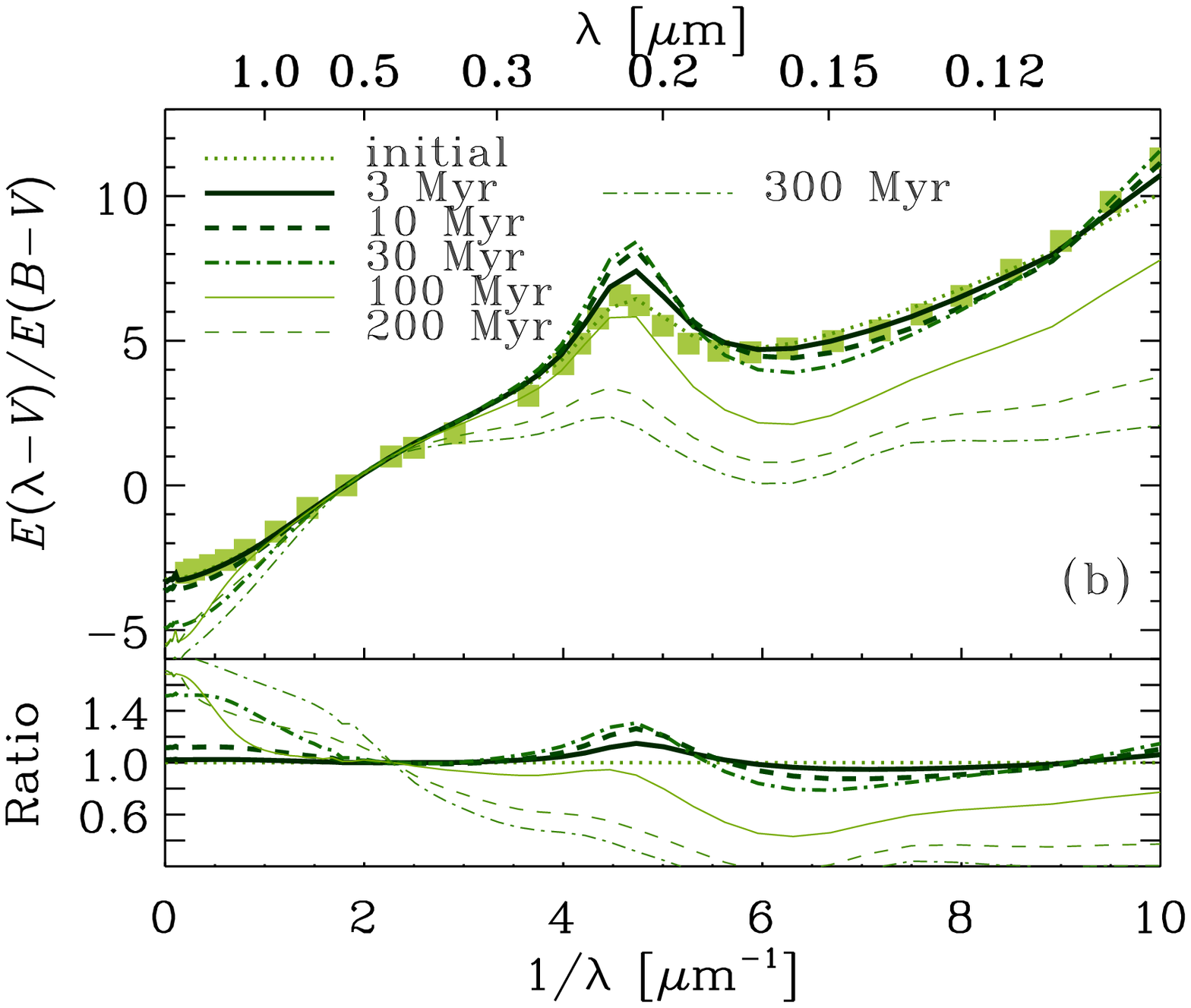}
\caption{
Same as Fig.\ \ref{fig:ext} but for the tuned model.
The values of $R_V$ are 3.26, 3.33, 3.63, 4.94, 5.48, 5.56,
and 6.40 at $t=0$, 3, 10, 30, 100, 200, and 300 Myr, respectively.
\label{fig:ext_tuned}}
\end{figure*}

Based on the comparisons with observational data in
the above, we propose the following tuning for the
model. Since the coagulation of carbonaceous dust
proves to be too efficient, we decrease the coagulation
efficiency by a factor of 2 by adopting
$S_\mathrm{coag}=0.5$ for carbonaceous dust, while
we keep using $S_\mathrm{coag}=1$ for silicate.
For silicate, we get rid of the coagulation threshold;
that is, silicate grains are assumed to coagulate
whenever they collide. Indeed, if we consider fluffiness
of coagulated grains, the structure can efficiently
absorb the energy in collision and large grain can be
formed after sticking \citep{ormel09}. We do not change
the accretion
efficiency ($S_\mathrm{acc}=0.3$) for both silicate and
carbonaceous dust to minimize the fine
tuning. This is called `tuned model', while
we call the original model with $S_\mathrm{coag}=1$
and the coagulation threshold `original model
without tuning'. As shown later, such a minimum tuning of
the original model is enough to explain the variation of
main extinction curve features. We do not tune
accretion since the final result is only sensitive to
the relative efficiencies of accretion and coagulation
(so we only tune coagulation to avoid the degeneracy).

The evolution of grain size distribution for the tuned
model is shown in Fig.\ \ref{fig:size_tuned}. For
silicate, compared with the dust distributions in
the original model without tuning (Fig.\ \ref{fig:size}),
the grains grow further, even beyond 0.2 $\micron$
at $>100$ Myr. This is simply because we removed the
coagulation threshold so that the grains can coagulate
at any velocities. For carbonaceous dust, the effect of
tuning slightly increases the abundance of small grains
($\la 300$ \AA) that can
contribute to the carbon bump, although the change is not
so drastic as
for silicate. Nevertheless, we will show later that such a
small change has a significant impact on the
strength of carbon bump.

Fig.\ \ref{fig:ext_tuned} shows the extinction curves
for the tuned model. $A_V/N_\mathrm{H}$ decreases
at $>30$~Myr when $R_V\ga 5$. This is consistent with
the observed trend that $A_V/N_\mathrm{H}$ decreases
as $R_V$ increases
(Fig.\ \ref{fig:ext_obs}). This decrease is
particularly driven by coagulation of silicate
grains. The tuned model also keeps the bump strong
even at later epochs, which is also consistent
with the observed extinction curves.
The more prominent
carbon bump in the tuned model than in the original
model without tuning is due to slower coagulation
of carbonaceous dust.
For $E(\lambda -V)/E(B-V)$, the variation of
observed UV extinction curves in Fig.\ \ref{fig:ext_obs},
compared with that of theoretical ones in
Fig.\ \ref{fig:ext_tuned},
can be interpreted as the flattening at $\ga 30$ Myr
by coagulation.

\subsection{Relation between depletion and $R_V$}
\label{subsec:depletion}

\begin{figure}
\includegraphics[width=0.45\textwidth]{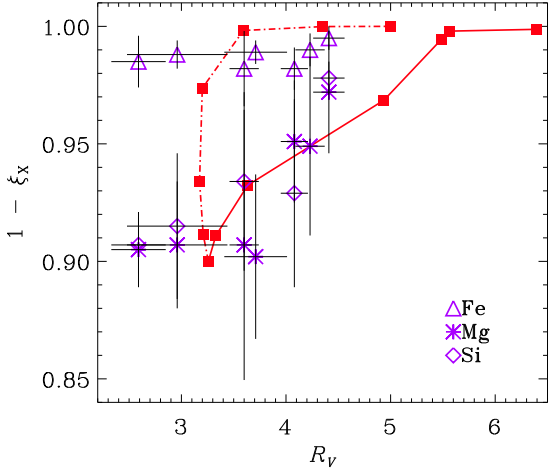}
\caption{Relation between $1-\xi_\mathrm{X}$
(the fraction of element of X in dust phase) and
$R_V$ (in the extinction curve) at 0, 3, 10, 30, 100, 200,
and 300 Myr for increasing $1-\xi_\mathrm{X}$.
The filled squares connected with the solid and dashed
lines show the theoretical results for the tuned model
and the original model without tuning. The triangles,
asterisks, and diamonds are the data taken from VH10
for X = Fe, Mg, and Si (in the models we adopt Si for
the tracer element).
\label{fig:dep_rv}}
\end{figure}

VH10 use $1-\xi_\mathrm{X}$
(in their notation $1-D_\mathrm{X}$)
to quantify the fraction of element X condensed
into dust grains. In particular, they show that
$1-\xi_\mathrm{X}$ has a correlation with
the shape of extinction curve. As mentioned in
Section \ref{subsec:ext}, we adopt $R_V$ as a
representative quantity for the extinction curve
slope. VH10 focus on silicate.
Since we adopted Si as a key element of silicate,
we concentrate on the relation between
$1-\xi_\mathrm{Si}$ and $R_V$ (note that
$R_V$ is also affected by carbonaceous dust
as well as silicate).
In Fig.\ \ref{fig:dep_rv}, we compare the
relations calculated by the models with the
observational data (we also plot the observed
depletions of Fe and Mg as references).

First, we discuss the original model without
tuning shown by the dashed line in
Fig.\ \ref{fig:dep_rv}. The slight decrease of $R_V$
at $\la 10$ Myr is due to accretion. Accretion
increases $1-\xi_\mathrm{X}$ rapidly
in 30 Myr. After 30 Myr, coagulation increases $R_V$ as
the extinction curve becomes flat. As a result,
$1-\xi_\mathrm{X}$ increases
(and $R_V$ decreases slightly) before $R_V$ increases.
This behaviour on the $(1-\xi_\mathrm{X})$--$R_V$
diagram does not explain the observed positive
correlation between
these two quantities in Fig.\ \ref{fig:dep_rv}.
The discrepancy implies that accretion
proceeds too quickly compared with coagulation
in the original model without tuning.

Next, we examine the tuned model. Note that the
tuning is meant to reproduce the observed trend
in extinction curves so it is not obvious
if it reproduces the observed relation
between depletion and $R_V$. As is clear from
Fig.\ \ref{fig:dep_rv}, the tuned model is consistent
with the observed trend. This is because of
more efficient coagulation of silicate, which pushes
$R_V$ to larger values at earlier epochs.
Therefore, the tuned model reproduces the
observed relation between depletion and $R_V$.

\section{Correlations among Various Extinction Properties}
\label{sec:fitzpatrick}

Now we discuss the models in the context of
the largest sample of Milky Way extinction curves
taken by \citet{fitzpatrick07}. \citet{cardelli89}
suggest that $R_V$ correlates with UV slope and
carbon bump strength.
More recently, \citet{fitzpatrick07} show that
the correlation is not significant for the bulk of
the sample with
$2.4<R_V<3.6$ but that there is a slight trend if
data points with larger $R_V$ values are included
(see also below).
We examine if these weak correlations reflect
the evolution of grain size distribution by
accretion and coagulation. More precisely, since
the correlations are weak, we check if the
models are consistent with the observed range
of those parameters. As mentioned in
Section~\ref{subsec:ext}, we examine
$c_2$, $\Delta 1250$, $A_\mathrm{bump}$, and
$E_\mathrm{bump}$ as characteristic quantities
in the extinction curve.

\begin{figure*}
\includegraphics[width=0.45\textwidth]{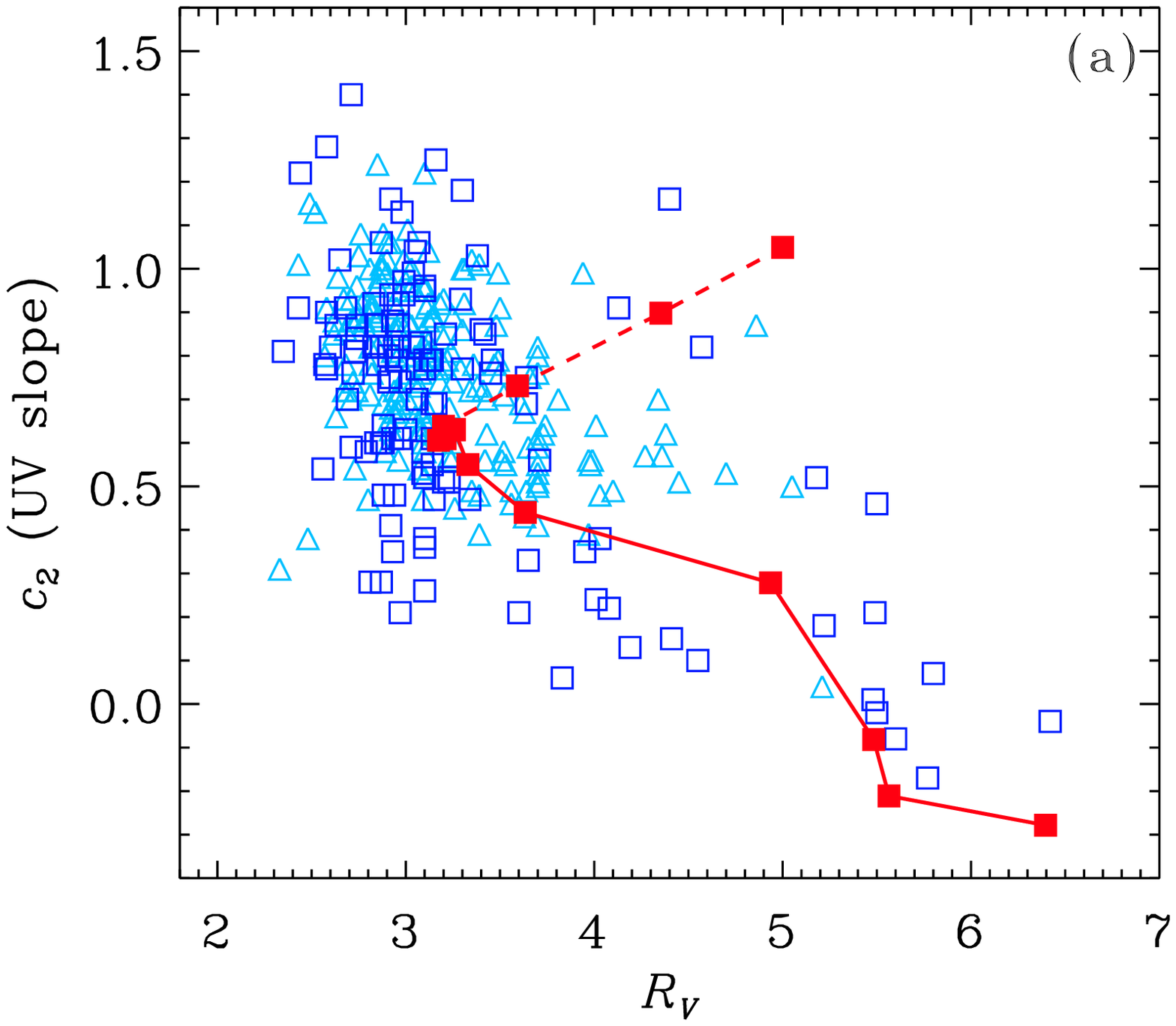}
\includegraphics[width=0.45\textwidth]{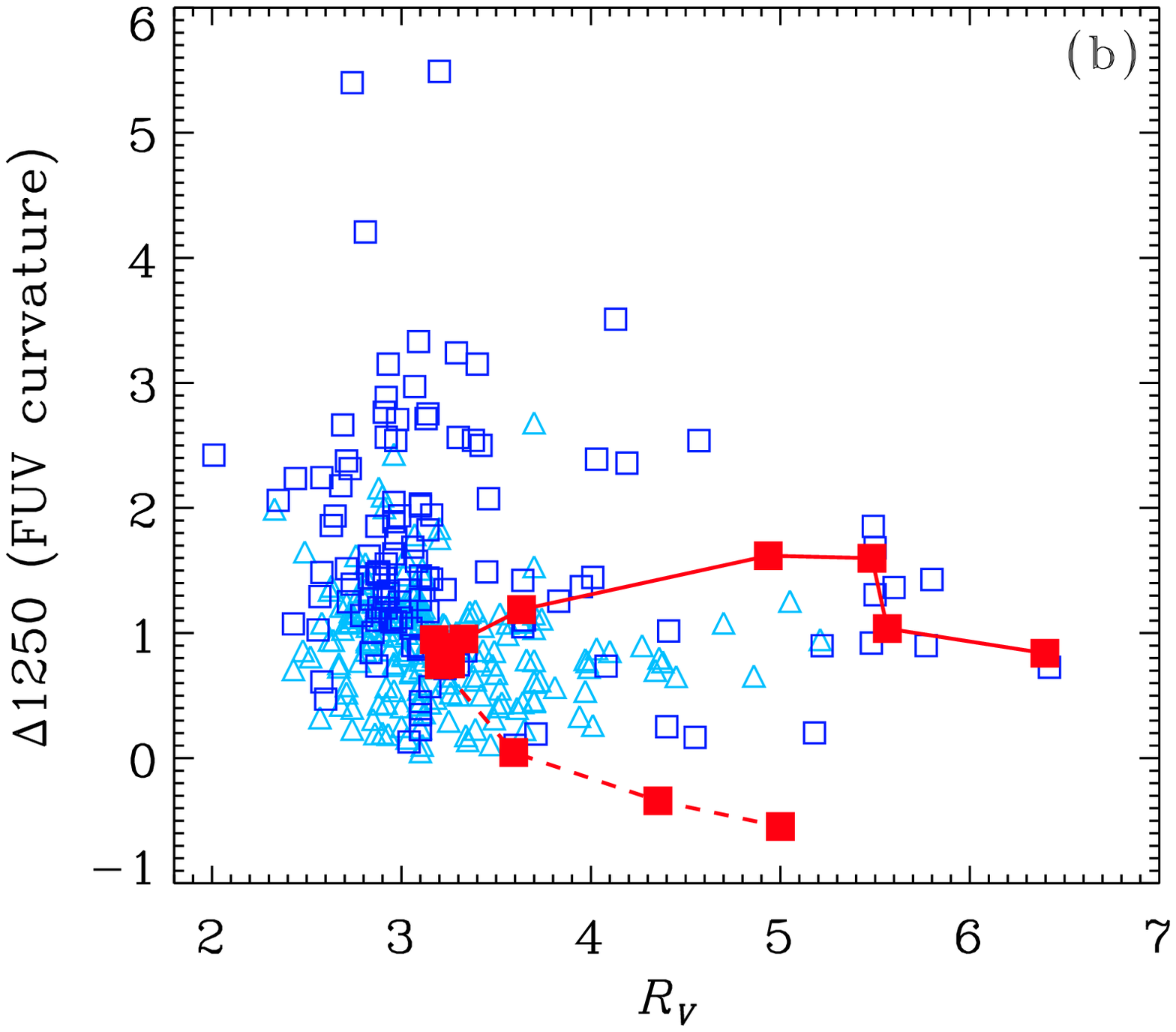}
\includegraphics[width=0.45\textwidth]{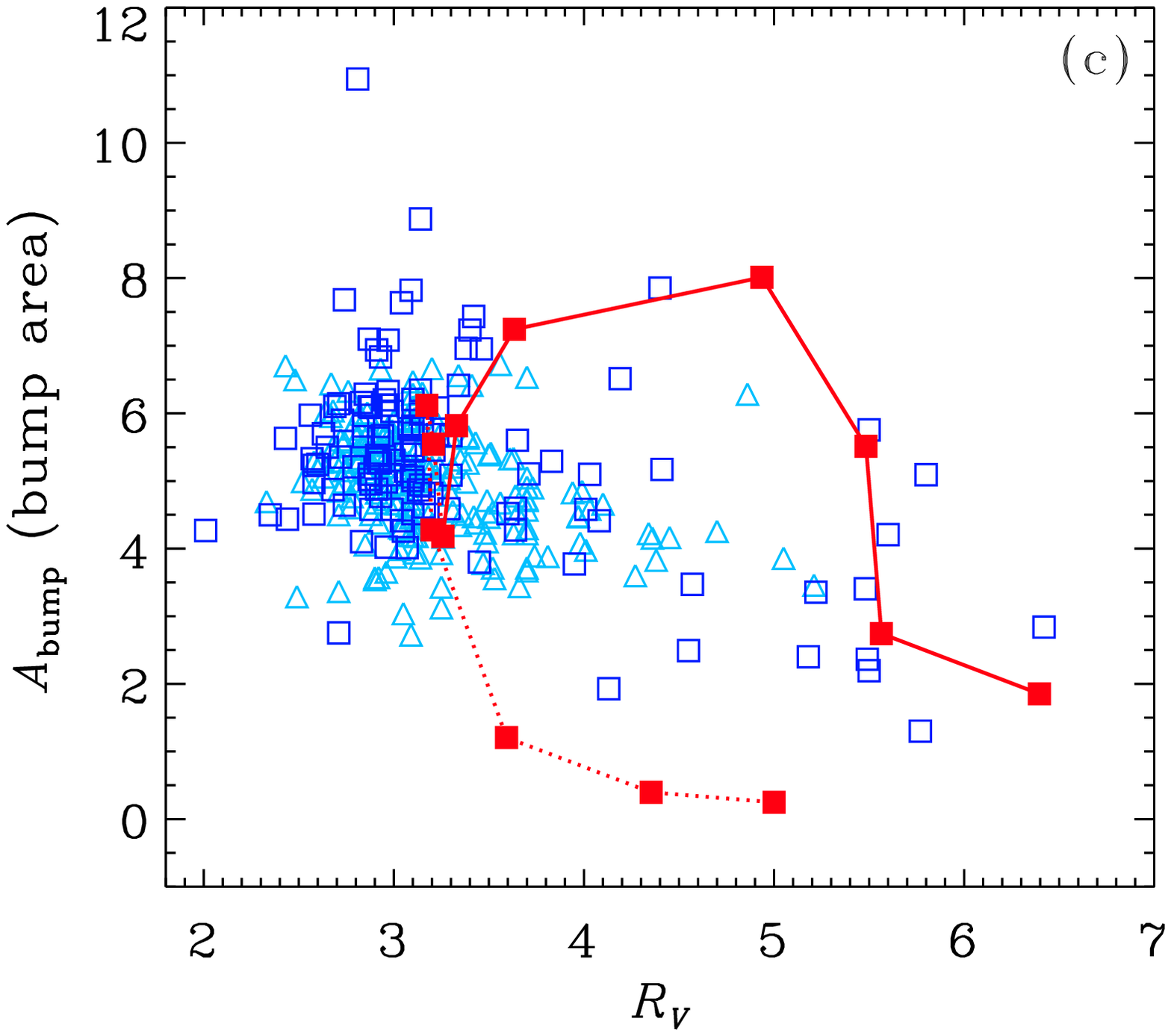}
\includegraphics[width=0.45\textwidth]{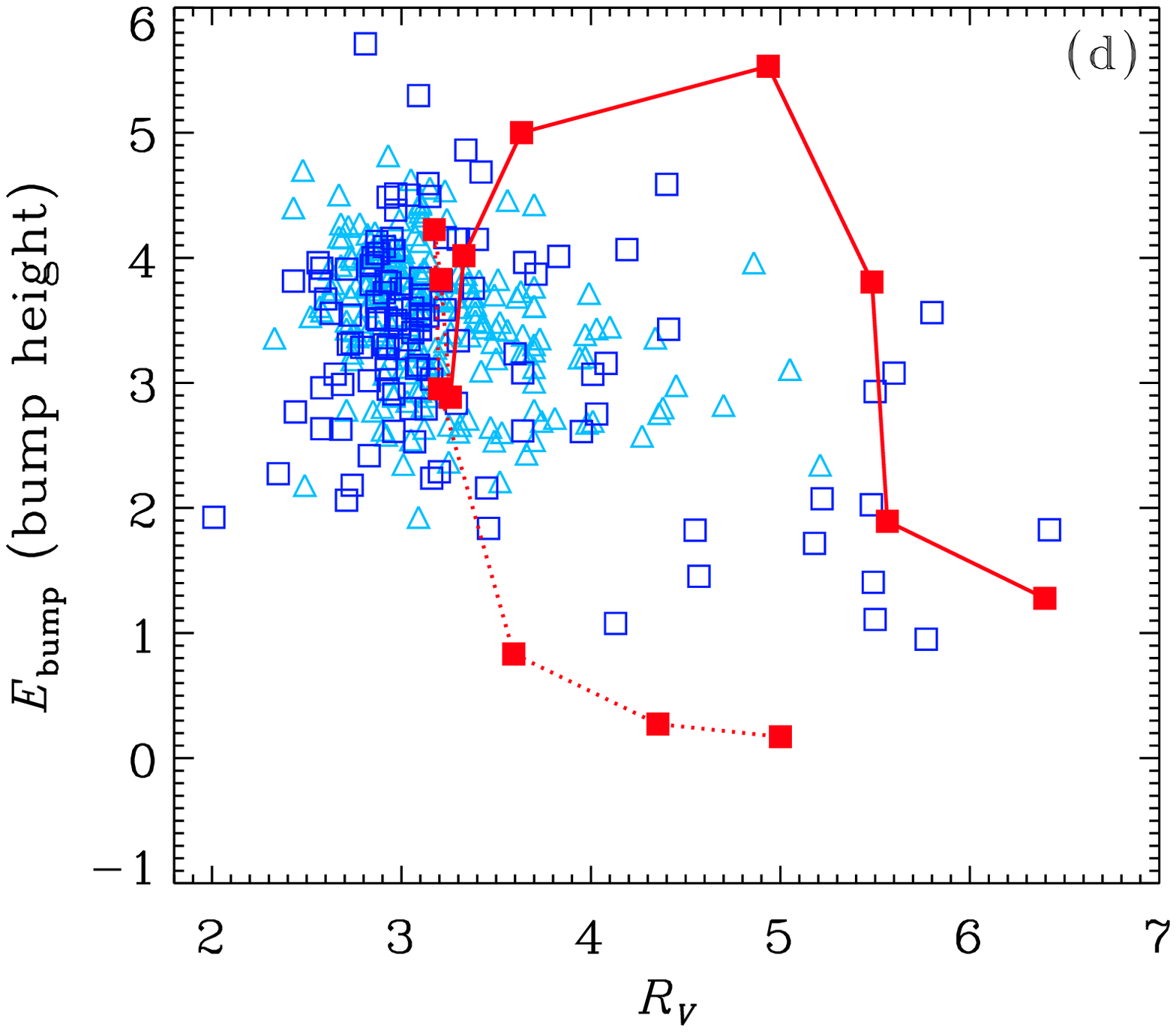}
\caption{Relation between $R_V$ and each of the characteristic
parameters of extinction curve
($c_2$, $\Delta 1250$, $A_\mathrm{bump}$, and
$E_\mathrm{bump}$ in Panels a, b, c, and d, respectively).
Our theoretical results are shown by
the filled squares connected with lines at
$t=0$, 3, 10, 30, 100, 200, and 300
Myr (the dashed and solid lines correspond to the original
model without tuning and the tuned model, respectively).
The evolution starts with the point with $R_V=3.2$.
The points for $t=3$, 10, and 30 Myr on the original
model without tuning
in Panels a and b are almost identical to the point at
$t=0$ Myr. The data points (open triangles and open
squares for stars
at $>1$ kpc and $<1$ kpc, respectively)
derived from
extinction curves in the Milky Way are taken from
\citet{fitzpatrick07}.
\label{fig:ext_para}}
\end{figure*}

In Fig.\ \ref{fig:ext_para}, we show the relation
between each of these parameters ($c_2$, $\Delta 1250$,
$A_\mathrm{bump}$, or $E_\mathrm{bump}$) and $R_V$.
For the observational data, the stars at $>1$ kpc
(called distant sample) and $<1$ kpc (called nearby sample)
are shown in different symbols. For all those parameters,
the dispersion is
smaller for the distant sample than for the nearby sample.
This indicates that the peculiarity in each individual
line of sight is reflected more in the nearby sample while
the peculiarity is `averaged' for the distant sample.
Even for the distant sample, some trends still remain:
$c_2$ and the carbon bump strength both have negative
correlations with $R_V$, although the correlations are
weak. The UV curvature has a weak positive trend with
$R_V$ for $R_V>3.5$, although there also seems to be
an overall negative trend if we also include the data
with smaller $R_V$. Because of the large variety in
individual lines of sight and the weakness of the trend,
we only use the relations in Fig.\ \ref{fig:ext_para} as
a reference for our models and do not attempt a fit to
the data. Although the initial condition may vary,
we simply adopt the same models used above, that is,
the original model without tuning and the tuned model
with the initial condition described in
Section \ref{subsec:initial}, in order to focus on
the trend produced by accretion and coagulation.

For the UV slope, $c_2$, the original model without
tuning predicts increasing $c_2$ (i.e.\ steepening of
the UV extinction curve) for increasing $R_V$ although
as shown in Fig.\ \ref{fig:ext}, coagulation tends to
flatten the overall UV extinction curve. As mentioned
in Section~\ref{subsec:ext},
$c_2$ reflects the slope around the carbon bump.
We find that $E(\lambda -V)/E(B-V)$ decreases around
$1/\lambda\sim 4~\micron^{-1}$ while it changes little
around $1/\lambda\sim 6~\micron^{-1}$ in the original
model without tuning. The decrease
around $1/\lambda\sim 4~\micron^{-1}$
is due to the flattening of carbon extinction
curve. On the other hand, silicate extinction changes
insignificantly since the silicate abundance at
$a>0.03~\micron$, where the effects on $c_2$
is the largest, is little affected by coagulation
(because their velocities are larger than the
coagulation threshold),
so the extinction around $1/\lambda\sim 6~\micron^{-1}$,
which is dominated by silicate, does not change much.
Thus, the
extinction curve seems to steepen in terms of
$c_2$. The tuned model seems to better reproduce the
decreasing trend of $c_2$ for increasing $R_V$.

For the FUV curvature, $\Delta 1250$, we reproduce
the observed trend of decreasing $\Delta 1250$ for
increasing $R_V$ (Fig.\ \ref{fig:ext_para}b) with the
original model without tuning. However, after 100~Myr,
$\Delta 1250$ becomes negative, which reflects
the convex shape
at $1/\lambda > 6~\micron^{-1}$. No Milky Way
extinction curve shows negative $\Delta 1250$.
The tuned model is rather near to the observational
data points.

Also for the bump strengths indicated by
$A_\mathrm{bump}$ and $E_\mathrm{bump}$, the original
model without tuning can explain the trend of
a less prominent bump for a larger $R_V$. However,
the bump weakens too much: Fig.\ \ref{fig:ext} shows
that the bump eventually disappears after 100 Myr, which
is not consistent with the fact that an extinction curve
without carbon (2175-\AA) bump is never observed in
the Milky Way. The tuned model keeps the bump strong
as we observe in Fig.\ \ref{fig:ext_tuned}.
The more prominent
carbon bump in the tuned model than in the original
model without tuning is due to slower coagulation of
carbonaceous dust.
The enhancement of the bump by accretion at $t\la 10$ Myr,
which may
reproduce the upward scatter in the bump strengths,
is also worth noting. However, this enhancement
causes too large bump strengths around $R_V\sim 5$,
although it is interesting that the data points
around $R_V>5.5$ are again reproduced by the
tuned model. 
The implication is that the scatter
of the observational data can reflect the
strong dependence on coagulation efficiency,
which may be changed by the detailed material
properties of dust.
We emphasize that only a factor of two variation
in $S_\mathrm{coag}$ (0.5 and 1) easily cover the
observed range of bump strength.

\section{Discussion}\label{sec:discussion}

\subsection{Implications of the tuned model}

In the tuned model, the coagulation threshold of silicate
is removed,
since the stop of coagulation at small sizes
prevents the extinction curve from becoming
flat enough to explain the observed UV extinction
curves for large $R_V$. The coagulation efficiency
of carbonaceous dust is reduced (only) by a factor
of 2 (i.e.\ we adopted $S_\mathrm{coag}=0.5$) to
keep enough abundance of small carbonaceous
grains contributing to the carbon bump. Those
changes are already enough
to reproduce the observational trend in the shape
of extinction curves for increasing $R_V$.
The UV slope, the FUV curvature and the carbon
bump strength have weak correlations with $R_V$.
The original model without tuning and the tuned model
predict very different tracks in the relations between
$R_V$ and each of those quantities. This implies that
the large scatters of those quantities  in the observational
data is due to the sensitive response to the
material properties (in our models, the coagulation
efficiency).

\subsection{Lifetime of molecular clouds}
\label{subsec:lifetime}

In this paper, we adopted the duration of accretion
and coagulation up to 300 Myr. This may be long
as a lifetime of molecular clouds
\citep[e.g.][]{lada10}, but \citet{koda09}
suggest a possibility that molecular clouds are sustained
over the circular
time-scale in a spiral galaxy ($\sim 100$ Myr).
Thus, it is worth investigating such a long
duration as a few hundreds of mega-years for
accretion and coagulation in molecular clouds.

The duration $t$ is also degenerate with
$n_\mathrm{H}Z$ as mentioned in
Section \ref{sec:model} in such a way that
the same value of $tn_\mathrm{H}Z$ returns the
same result. For example, if we adopt
$n_\mathrm{H}=10^4$ cm$^{-3}$ instead of
$10^3$ cm$^{-3}$, $R_V\sim 5$ is reached
at $t\sim 10$ Myr. However, the tracks on the
diagrams shown in Figs.\ \ref{fig:dep_rv}
and \ref{fig:ext_para} do not change
even if we change $n_\mathrm{H}Z$. In other
words, our predictions on the theoretical
tracks in these diagram are robust
against the variation of gas density.

\section{Conclusion}\label{sec:conclusion}

We have examined the effects of two major growth
mechanisms of dust grains, accretion and coagulation,
on the extinction curve. First we examined the evolution
of extinction curve by using the prescription for
accretion and coagulation with commonly used material
parameters. The observational Milky Way
extinction curves are not reproduced in the following
two respects: (i) The
extinction normalized to the hydrogen column density
($A_V/N_\mathrm{H}$) does not decrease around
$1/\lambda\sim 6$--8 $\micron^{-1}$ in the model even
for large $R_V (> 4)$.
(ii) The carbon bump is too small when $R_V>4$.
These two discrepancies are resolved by the following
prescriptions (tuning):
more efficient coagulation for silicate by removing the
coagulation threshold velocity, and less efficient coagulation
for carbonaceous dust by a factor of 2. This
`tuned' model also reproduces the trend between
depletion and $R_V$, which implies that the
time-scale of coagulation relative to that of accretion is
appropriate. On the other hand, the original model
without tuning fails to reproduce the observed trend
between depletion and $R_V$ since too quick accretion
relative to coagulation increases depletion even
before coagulation increases $R_V$.

We have also examined the relation between $R_V$ and
each of the following extinction curve features:
the UV slope, the FUV curvature, and the carbon bump
strength. The correlation between the UV slope and $R_V$,
which is the strongest among those three correlations,
is well reproduced by the tuned model. For the FUV curvature
and the carbon bump, the observational data are located
between the original model without tuning and the tuned
model, which implies that the scatters in the observational data
can be attributed to the sensitive variation to the dust properties
(coagulation efficiency in our models). This also means that
the variation of extinction curves in the Milky Way, especially
at $R_V>3$, can be interpreted by the variation of grain
size distribution by accretion and coagulation. For more
complete understanding, especially at $R_V<3$, other
dust processing mechanisms such as shattering
should be included.

\section*{Acknowledgments}

%%We thank K. Sakamoto for his continuous help
%%for the SMA observation and the data analysis.
%%We are grateful to the
%%anonymous referee for useful comments that improved
%%this paper very much.
HH thanks the support from
NSC grant NSC102-2119-M-001-006-MY3, and
NVV acknowledges the support from the Grant RFBR 13-02- 00138a.

%%\appendix

%%\section{Data for the radio--FIR relation}
%%\label{app:radio_FIR}

\bsp

\label{lastpage}


\begin{thebibliography}{}
\bibitem[\protect\citeauthoryear{Asano et al.}{2013a}]{asano13}
    Asano, R., Takeuchi, T. T., Hirashita, H., \& Nozawa, T. 2013a,
    MNRAS, 432, 637
\bibitem[\protect\citeauthoryear{Asano et al.}{2013b}]{asano12}
    Asano, R., Takeuchi, T. T., Hirashita, H., \& Inoue, A. K.
    2013b, Earth Planets Space, 65, 213
\bibitem[\protect\citeauthoryear{Bianchi \& Schneider}{2007}]{bianchi07}
    Bianchi, R., \& Schneider, R. 2007, MNRAS, 378, 973
\bibitem[\protect\citeauthoryear{Bohren \& Huffman}{1983}]{bohren83}
    Bohren C. F., Huffman D. R., 1983, Absorption and Scattering of Light
    by Small Particles. Wiley, New York
\bibitem[\protect\citeauthoryear{Cardelli, Clayton, \& Mathis}{Cardelli et al.}{1989}]{cardelli89}
    Cardelli J. A., Clayton G. C., Mathis J. S., 1989, ApJ, 345, 245
\bibitem[\protect\citeauthoryear{Cazaux \& Tielens}{2004}]{cazaux04}
    Cazaux, S., \& Tielens, A. G. G. M. 2004, ApJ, 604, 222
\bibitem[\protect\citeauthoryear{Chokshi, Tielens, \& Hollenbach}{1993}]{chokshi93}
    Chokshi, A., Tielens, A. G. G. M., \& Hollenbach, D. 1993, ApJ,
    407, 806
\bibitem[\protect\citeauthoryear{Cox}{2000}]{cox00}
    Cox, A. N. 2000, Allen's Astrophysical Quantities, 4th ed., Springer,
    New York
\bibitem[\protect\citeauthoryear{D\'{e}sert, Boulanger, \& Puget}{1990}]{desert90}
    D\'{e}sert, F.-X., Boulanger, F., \& Puget, J. L. 1990,
    A\&A, 237, 215
\bibitem[\protect\citeauthoryear{Dominik \& Tielens}{1997}]{dominik97}
    Dominik, C., \& Tielens, A. G. G. M. 1997, ApJ, 480, 647
\bibitem[\protect\citeauthoryear{Draine}{2009}]{draine09}
    Draine B. T., 2009, in Henning Th., Gr\"{u}n E., Steinacker J., eds,
    ASP Conf.\ Ser.\ 414, Cosmic Dust -- Near and Far.
    Astron.\ Soc.\ Pac., San Francisco, p.\ 453
\bibitem[\protect\citeauthoryear{Draine \& Lee}{1984}]{draine84}
    Draine, B. T., \& Lee, H. M. 1984, ApJ, 285, 89
\bibitem[\protect\citeauthoryear{Draine \& Sutin}{1987}]{draine87}
    Draine, B. T., \& Sutin, B. 1987, ApJ, 320, 803
\bibitem[\protect\citeauthoryear{Dwek}{1998}]{dwek98}
    Dwek E., 1998, ApJ, 501, 643
\bibitem[\protect\citeauthoryear{Fitzpatrick \& Massa}{2007}]{fitzpatrick07}
    Fitzpatrick, E. L., \& Massa, D. 2007, ApJ, 663, 320
\bibitem[\protect\citeauthoryear{Hirashita}{2000}]{hirashita00}
    Hirashita H., 2000, PASJ, 52, 585
\bibitem[\protect\citeauthoryear{Hirashita}{2012}]{hirashita12}
    Hirashita, H. 2012, MNRAS, 422, 1263 (H12)
\bibitem[\protect\citeauthoryear{Hirashita \& Li}{2013}]{hirashita13}
    Hirashita, H., \& Li, Z.-Y. 2013, MNRAS, 434, L70
\bibitem[\protect\citeauthoryear{Hirashita \& Yan}{2009}]{hirashita09}
    Hirashita, H., \& Yan, H. 2009, MNRAS, 394, 1061
\bibitem[Inoue(2011)]{inoue11} Inoue, A. K. 2011,
    Earth, Planets Space, 63, 1027
\bibitem[\protect\citeauthoryear{Jones \& Nuth}{2011}]{jones11}
    Jones, A. P., \& Nuth, J. A., III 2011, A\&A, 530, A44
\bibitem[\protect\citeauthoryear{Jones, Tielens, \& Hollenbach}{Jones et al.}{1996}]{jones96}
    Jones, A. P., Tielens, A. G. G. M., \& Hollenbach, D. J. 1996, ApJ,
    469, 740
\bibitem[\protect\citeauthoryear{Jones et al.}{1994}]{jones94}
    Jones, A. P., Tielens, A. G. G. M., Hollenbach, D. J., \&
    McKee, C. F. 1994, ApJ, 433, 797
\bibitem[\protect\citeauthoryear{Koda et al.}{2009}]{koda09}
    Koda, J., et al.\ 2009, ApJ, 700, L132
\bibitem[\protect\citeauthoryear{Lada, Lombardi, \& Alves}{2010}]{lada10}
    Lada, C. J., Lombardi, M., \& Alves, J. F. 2010, ApJ, 724, 687
\bibitem[\protect\citeauthoryear{Li \& Draine}{2001}]{li01}
    Li, A., \& Draine, B. T. 2001, ApJ, 554, 778
%%\bibitem[\protect\citeauthoryear{Mathis, Rumpl, \& Nordsieck}{Mathis et al.}{1977}]{mathis77}
%%    Mathis, J. S., Rumpl, W., \& Nordsieck, K. H. 1977, ApJ, 217, 425
%%\bibitem[\protect\citeauthoryear{Mattsson}{2011}]{mattsson11}
%%    Mattsson, L. 2011, MNRAS, 414, 781
\bibitem[\protect\citeauthoryear{McKee}{1989}]{mckee89}
    McKee, C. F. 1989, in Allamandola L. J. \& Tielens A. G. G. M. eds.,
    IAU Symp.\ 135, Interstellar Dust, Kluwer, Dordrecht, 431
\bibitem[Nozawa et al.(2007)]{nozawa07} Nozawa, T., Kozasa, T.,
    Habe, A., Dwek, E., Umeda, H., Tominaga, N., Maeda, K., \&
    Nomoto, K. 2007, ApJ, 666, 955
\bibitem[\protect\citeauthoryear{O'Donnell \& Mathis}{1997}]{odonnell97}
    O'Donnell, J. E., \& Mathis, J. S. 1997, ApJ, 479, 806
\bibitem[\protect\citeauthoryear{Ormel et al.}{2009}]{ormel09}
    Ormel, C. W., Paszun, D., Dominik, C., \& Tielens, A. G. G. M. 2009,
    A\&A, 502, 845
%%\bibitem[\protect\citeauthoryear{Osterbrock}{1989}]{osterbrock89}
%%    Osterbrock, D. E. 1989,
%%    Astrophysics of Gaseous Nebulae and Active Galactic Nuclei
%%    (Mill Valley: University Science Books)
\bibitem[\protect\citeauthoryear{Pei}{1992}]{pei92}
    Pei Y. C., 1992, ApJ, 395, 130
\bibitem[\protect\citeauthoryear{Pipino et al.}{2011}]{pipino11}
    Pipino A., Fan X. L., Matteucci, F., Calura F., Silva L., Granato G.,
    Maiolino R., 2011, A\&A, 525, A61
%%\bibitem[Rybicki \& Lightman(1979)]{rybicki79} Rybicki, G. B., \&
%%    Lightman, A. P. 1979, Radiative Processes in Astrophysics
%%    (New York: Wiley)
\bibitem[\protect\citeauthoryear{Savage \& Sembach}{1996}]{savage96}
    Savage B. D., Sembach K. R., 1996, ARA\&A, 34, 279
%%\bibitem[\protect\citeauthoryear{Schmidt \& Boller}{1993}]{schmidt93}
%%    Schmidt, K.-H., \& Boller, T. 1993, Astron.\ Nachr., 314, 361
%%\bibitem[\protect\citeauthoryear{Spitzer}{1978}]{spitzer78}
%%    Spitzer, L. 1978, Physical
%%    Processes in the Interstellar Medium (New York: Wiley)
\bibitem[\protect\citeauthoryear{Stepnik et al.}{2003}]{stepnik03}
    Stepnik B. et al., 2003, A\&A, 398, 551
%%\bibitem[\protect\citeauthoryear{Takeuchi et al.}{2005}]{takeuchi05}
%%    Takeuchi, T. T.,
%%    Ishii, T. T., Nozawa, T., Kozasa, T., \& Hirashita, H. 2005,
%%    MNRAS, 362, 592
\bibitem[\protect\citeauthoryear{Valiante et al.}{2011}]{valiante11}
    Valiante, R., Schneider, R., Salvadori, S., \& Bianchi, S.
    2011, MNRAS, 416, 1916
\bibitem[\protect\citeauthoryear{Voshchinnikov}{2012}]{voshchinnikov12}
    Voshchinnikov, N. V. 2012,
    Journal of Quantitative Spectroscopy \& Radiative Transfer, 113, 2334
\bibitem[\protect\citeauthoryear{Voshchinnikov \& Henning}{2010}]{voshchinnikov10}
    Voshchinnikov, N. V., \& Henning, Th.\ 2010, A\&A, 517, A45 (VH10)
\bibitem[\protect\citeauthoryear{Weingartner \& Draine}{1999}]{weingartner99}
    Weingartner, J. C., \& Draine, B. T. 1999, ApJ, 517, 292
\bibitem[\protect\citeauthoryear{Weingartner \& Draine}{2001}]{weingartner01}
    Weingartner, J. C., \& Draine, B. T. 2001, ApJ, 548, 296
\bibitem[\protect\citeauthoryear{Yamasawa et al.}{2011}]{yamasawa11}
    Yamasawa, D., Habe, A., Kozasa, T., Nozawa, T., Hirashita, H.,
    Umeda, H., \& Nomoto, K. 2011, ApJ, 735, 44
\bibitem[\protect\citeauthoryear{Yan, Lazarian \& Draine}{2004}]{yan04}
    Yan H., Lazarian A., Draine B. T., 2004, ApJ, 616, 895
\bibitem[\protect\citeauthoryear{Yasuda \& Kozasa}{2012}]{yasuda12}
    Yasuda, Y., \& Kozasa, T. 2012, ApJ, 745, 159
\bibitem[\protect\citeauthoryear{Zhukovska, Gail, \& Trieloff}{2008}]{zhukovska08}
    Zhukovska S., Gail H.-P., Trieloff M., 2008, A\&A, 479, 453
\end{thebibliography}
\end{document}